\documentclass[twoside,twocolumn,english,aps,amsmath,amssymb,reprint,pra,superscriptaddress]{revtex4-1}
\usepackage[T1]{fontenc}
\usepackage[latin9]{inputenc}
\usepackage{color}
\usepackage{amsmath}
\usepackage{graphicx}

\makeatletter
\usepackage{babel}

\makeatother

\usepackage{babel}
\begin{document}

\title{Quantum multiplexing }

\author{Nicol$\acute{{\rm {\bf o}}}$ Lo Piparo}
\email{nicopale@gmail.com}

\affiliation{National Institute of Informatics, 2-1-2 Hitotsubashi, Chiyoda, Tokyo
101-0003, Japan.}

\author{William J. Munro}

\affiliation{NTT Basic Research Laboratories \& NTT Research Center for Theoretical
Quantum Physics, NTT Corporation, 3-1 Morinosato-Wakamiya, Atsugi,
Kanagawa, 243-0198, Japan.}

\affiliation{National Institute of Informatics, 2-1-2 Hitotsubashi, Chiyoda, Tokyo
101-0003, Japan.}

\author{Kae Nemoto}

\affiliation{National Institute of Informatics, 2-1-2 Hitotsubashi, Chiyoda, Tokyo
101-0003, Japan.}
\begin{abstract}
Distributing entangled pairs is a fundamental operation required for
many quantum information science and technology tasks. In a general
entanglement distribution scheme, a photonic pulse is used to entangle
a pair of remote quantum memories. Most applications require multiple
entangled pairs between remote users, which in turn necessitates several
photonic pulses (single photons) being sent through the channel connecting
those users. Here we present an entanglement distribution scheme using
only a single photonic pulse to entangle an arbitrary number of remote
quantum memories. As a consequence the spatial temporal resources
are dramatically reduced. We show how this approach can be simultaneously
combined with an entanglement purification protocol to generate even
higher fidelity entangled pairs. The combined approach is faster to
generate those high quality pairs and requires less resources in terms
of both matter qubits and photons consumed. To estimate the efficiency
of our scheme we derive a normalized rate taking into account the
raw rate at which the users can generate purified entangled pairs
divided by the total resources used. We compare the efficiency of
our system with the Deutsch protocol in which the entangled pairs
have been created in a traditional way. Our scheme outperforms this
approach both in terms of generation rate and resources required.
Finally we show how our approach can be extended to more general error
correction and detection schemes with higher normalized generation
rates naturally occurring. 
\end{abstract}
\maketitle

\section{Introduction}

It has long been known that the principles of quantum mechanics will
allow new technologies to be developed bringing significant performance
enhancements or the potential for new capabilities yet unrealized
with conventional technology \cite{feynman,simon,fDiVincenzo,Qtech}.
Such technologies can be broadly categorized into a number of groups
including quantum sensing and imaging \cite{QSensing,QImaging2,QImaging3},
quantum communication \cite{QComm,Quantumcomm,QKD01,QKD02,QKD03,QC_Bill}
and quantum computation \cite{Qcomp1,Qcomp2,Quantum_comp2,Quantum_comp3,Quantum_comp4,QComp_Bill}.
Many of these technologies are non-local in nature and require shared
entanglement between the remote users. Traditional communication applications
including quantum cryptography \cite{QC1,QC2,QC3,QC4} and quantum
teleportation \cite{tel_1,tel_2,Qtel2,Qtel3,Qtel4} are based on creating
entangled pairs between two parties (Alice and Bob). These pairs can
either be directly used \cite{Qentang,Qentang2,Qentang3} or stored
in quantum memories \cite{QMentanglement,QMentanglement2,QMentanglement3,QMentanglement4,QMentanglement5}. 

In most entanglement generation schemes the entanglement creation
is mediated by single photons, which travels across a lossy channel
between Alice to Bob \cite{QC_Bill}. Many systems require that multiple
entangled pairs are created simultaneously, for instance, in the multiple-memory
configuration used for quantum repeaters \cite{multiple_mem_conf,MultiQM2}
and in conventional purification protocols \cite{QP1,QP2Deutsch,QP3Dur}.
The creation of multiple pairs can be quite challenging especially
when the parties are separated by large distances due to channel losses.
This can cause significant performance issues \cite{Loss,Loss2}. 

There are a number of mechanisms for creating entangled pairs between
quantum memories (QMs) \cite{QMentanglement,QMentanglement5,QMent_creation}.
Generally these are based on quantum emitters \cite{QMent_emitters1,NVemitter1},
absorbers \cite{QMent_ab1,NVabs1} or conditional transmitters/reflecters
\cite{QM_refl1,NVreflect} and operate in systems including ion traps
\cite{trapped_ions,ionTrap1}, trapped atoms \cite{trapped_at,trapAtoms2},
quantum dots \cite{qdots,Qdots_ent2,Qdots_ent3} and Nitrogen-Vacancy
(NV), Silicon-Vacancy (SiV) centers in diamond \cite{NVcenters1,SiV_center}.
Recent NV experiments have created remote entanglement (and even violated
Bell's inequalities) using the emitter based approach \cite{NVexp1,NVexp2}.
However the low collection efficiency means the probability of success
(rate) for entangling the remote NV centers is small \cite{NVexp1,NVexp2}.
Embedding the NV center into a cavity is the natural way to increase
this collection efficiency but it also opens the possibility for using
the conditional transmission/reflection approaches \cite{NVcavity1,DE}.
Such NV based conditional transmission/reflection approaches have
been proposed for tasks ranging from conventional measurement device
independent QKD \cite{DE,singleNVQKD,NVKae} to quantum networks \cite{NVKae,NVnetworks1,NVnetworks2}
and large scale quantum computers \cite{NVcomputing2,QM_refl1}. Further
these approaches offer the possibility of having a single photon interacting
with multiple NV centers. In such a way we could create multiple entangled
pairs by using only one photon, which could help overcome channel
losses issues and therefore increase the communication rates. 

As an alternative to sending multiple independent photons for creating
entangled pairs between Alice and Bob we can in principle use multiple
degrees of freedom (DOFs) \cite{DOF1,DOF2,DOF3,DOF_Bill} in a single
photon to achieve the same purpose. This has the potential advantage
that, if the photon successfully reaches Bob, then multiple entangled
pairs are generated in one instance. Further, the probability of success
for transmitting a single DOFs-photon through the channel is higher
than the probability associated with transmitting multiple independent
(non-DOFs) photons through the same channel. However if no photon
is successfully transmitted in the DOF case then no entangled pair
is generated. 

In this work we combine the\textbf{ }transmission/reflection approach
of an NV center embedded in a cavity with multiple DOFs encoding of
a photon. \textcolor{black}{We call this method ``quantum multiplexing''
(QMUXING) as one photon carries multiple DOFs for entangling multiple
independent quantum memories across a long distance communication
channel. This is in contrast to traditional communication multiplexing
(time-frequency division multiplexing) where multiple signals are
transmitted through the channel at the same time.} Initially we show
that, in order to create two entangled pairs by using this method,
the average waiting time is lower than using two independent photons.
In this way, the dephasing effects on the quantum memories are less
detrimental than that conventional entangling scheme and the number
of photons is less. This has a significant impact on the performance
of the entanglement generation scheme, especially in terms of spatial
and temporal resources. We can naturally extend this approach to create
many entangled quantum memory pairs by adding further DOFs onto the
photon. Such pairs could be used directly for QKD where lower quality
entangled states are acceptable, but they can also be resource to
generate extremely high fidelity pairs using quantum error detection
and correction codes \cite{Qcomp1} for quantum computation.

Entanglement purification \cite{QP1,QP2Deutsch,QP3Dur,Quantum_pur1}
is the simplest error detection mechanism \cite{devitt} that can
be used to create high fidelity Bell pairs from lower fidelity ones.
In such systems, it is an essential requirement that we create several
entangled pairs by using independent photons and have these available
at nearly the same time. We can apply our entangling method to the
Deutsch purification protocol \cite{QP2Deutsch} and show that we
obtain higher entangling rates with less number of resources. Furthermore,
the number of quantum memories can be still reduced if we perform
the local operations required for the purification protocol directly
on the extra DOFs of the photon \cite{private}. In this way, we then
derive a protocol (QMUXING protocol) for creating high fidelity pairs
based on the QMUXING entangling scheme with fewer quantum memories
than the ones used in a conventional purification protocol. We can
naturally extend this approach to other error detection and correction
protocols \cite{Err.corr1,Err.corr2,Err.corr3,Err.corr4,Err.corr5,Err.corr6,Err.corr7}. 

To evaluate the performance of our QMUXING protocol, we consider the
rate at which Alice and Bob create an high fidelity pair normalized
by the total number of resources used (modified by a variable cost
function to take into account the relative impact that they might
have on a practical implementation). We calculate the normalized rate
of our protocol and compare it to the rate of creating high fidelity
pairs through the Deutsch protocol in which its entangled pairs are
created in a traditional way. We show that our protocol simnifically
outperforms the other systems. A higher normalized rate is also obtained
in the case of applying our scheme to a conventional three-qubit error
correction protocol.

Our paper is structured as follows: In Sec. II, we describe the simplest
application of the QMUXING for entangling four quantum memories and
discuss its main advantages compared to a traditional entangling scheme.
These advantages are extended to the Deutsch purification protocol
optimized by the QMUXING scheme, as shown in Sec. III. In this Section
we further describe the QMUXING protocol where only three memories
are used. We then present in Sec. IV analytical expressions for the
normalized rate in order to estimate the efficiency of the QMUXING
protocol. In Sec. V, we calculate the ratio between the normalized
rate of the QMUXING protocol and the conventional one for different
values of the cost functions and different distillation rounds. We
extend our analysis to the case of a three-qubits quantum error correction
protocol and we compare it with our scheme. Finally in Sec. VI we
provide a concluding discussion.

\section{\label{sec:quantum_multi_scheme}The quantum multiplexing entangling
scheme }

Let us now describe the quantum multiplexing based entangling scheme
applied to two pairs of NV centers separated by a distance $L$ before
we generalize it to an arbitrary number of NV centers. For the sake
of simplicity, we will assume that Alice and Bob create entangled
pairs through the usual ``prepare and measure'' protocol \cite{Prepare_meas_prot},
in which a photon, sent by Alice, ideally travels across the channel
until it reaches Bob's side. Here, an entangling mechanism will entangle
Alice and Bob's pair, upon a successful measurement of the photon.
Our QMUXING entangling scheme has analogous advantages when it is
applied to other types of entangling schemes, such as the man-in-the-middle
protocol or when a photonic entanglement source is located between
the users \cite{QC1}.

\subsection{The four qubit QMUXING entangling scheme }

The main building block of the QMUXING entangling scheme \cite{DE,NVKae}
operates by having a polarized photon interact and become entangled
with an internal degree of freedom (spin for instance) of the quantum
memory. \textcolor{black}{In our case we are considering NV centers
as th quantum memories.} Under an appropriate magnetic field the NV
center is an effective two level system, where we can use the $|m_{S}=0\left\rangle \right.$
and $|m_{S}=+1\left\rangle \right.$ states of the ground state manifold
as the qubit. We label these states as $|g\left\rangle \right.$ and
$|e\left\rangle ,\right.$ respectively. Our protocol begins when
the NV center is initialized in a superposition of electronic spin
states: $|\psi\left\rangle _{in}=\right.|g\left\rangle +|e\left\rangle ,\right.\right.$
where we have omitted the normalization constant for sake of simplicity.
A $D(A)$ polarized photons is then sent to the cavity where it will
interact and become entangled with the NV center. The interaction
of a photon with an NV center results in the ideal case with a $\pi$
phase shift on it when the NV center is in the $|e\left\rangle \right.$
state and the photon is vertically polarized. 
\begin{figure*}
\noindent \begin{centering}
\includegraphics[scale=0.37]{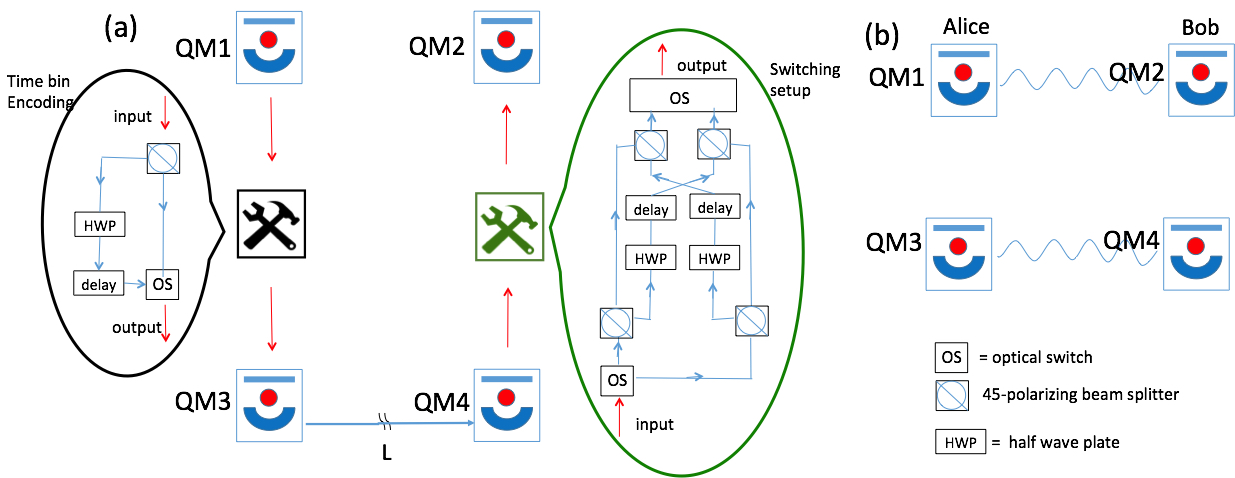}
\par\end{centering}
\caption{\label{fig:QM_4qbits}(a) Quantum multiplexing applied to the four
qubits entanglement distribution scheme (Alice and Bob have two qubits
each). A polarized photon is entangled with the electron spin states
of an NV center to create the state $|g_{1}\left\rangle |D\left\rangle \right.+|e_{1}\left\rangle |A\left\rangle .\right.\right.\right.$
The polarization encoded information is then transferred to a time
bin encoding on the photon through the \textquotedbl{}time bin encoding\textquotedbl{}
converter. The photon interacts with the second NV center and travels
across the channel to Bob. At Bob's side, the photon interacts with
an NV center followed by it passing through the \textquotedbl{}switching
setup\textquotedbl{} operation which swaps the polarization and time
bin degrees of freedom ($A_{\mathcal{S}}$ with the $D_{\mathcal{L}}$
mode). Finally, the photon interacts with the second NV center of
Bob and is measured. Upon a successful measurement, the state of the
four memories is projected into two maximally entangled states illustrated
in (b).}
\end{figure*}
In all the other cases no phase shift occurs. To illustrate this method,
we assume that Alice and Bob, separated by a distance $L,$ have two
NV centers each, as shown in Fig. \ref{fig:QM_4qbits}(a), respectively.
A $D-$ polarized photon interacts with the first NV center on Alice's
side giving \cite{DE} 

\noindent 
\begin{equation}
|D\left\rangle |\psi_{1}\left\rangle _{in}\right.\right.\rightarrow|g_{1}\left\rangle |D\left\rangle \right.+|e_{1}\left\rangle |A\left\rangle ,\right.\right.\right.
\end{equation}

\noindent The next step (as shown in Fig. \ref{fig:QM_4qbits}(a)
is to transfer the information encoded into the polarization DOF into
the time bin DOF through the \textquotedbl{}time bin encoding\textquotedbl{}
converter. \textcolor{black}{Our state is transformed to}

\noindent 
\begin{equation}
|g_{1}\left\rangle |D_{\mathcal{S}}\left\rangle \right.+|e_{1}\left\rangle |D_{\mathcal{L}}\left\rangle ,\right.\right.\right.
\end{equation}
where the subscripts $\mathcal{S}$ and $\mathcal{L}$ indicate the
short and long time-bins, respectively. The photon then interacts
with the second NV center of Alice (labelled QM3), giving: 

\noindent 
\begin{equation}
\begin{array}{c}
\begin{array}{c}
|g_{1}\left\rangle |g_{3}\left\rangle \right.|D_{\mathcal{S}}\left\rangle \right.+|g_{1}\left\rangle |e_{3}\left\rangle \right.|A_{\mathcal{S}}\left\rangle \right.\right.\right.\end{array}\\
\;\;\;\;\;\;\;\;\;+|e_{1}\left\rangle |g_{3}\left\rangle \right.|D_{\mathcal{L}}\left\rangle +\right.\right.|e_{1}\left\rangle |e_{3}\left\rangle \right.|A_{\mathcal{L}}\left\rangle .\right.\right.
\end{array}\label{eq:3rd_ent_step}
\end{equation}

\noindent The photon then travels through the optical fiber, where,
upon a successful transmission, it interacts with Bob's first qubit
(labelled QM4). While the photon's transmission through the channel
has a probabilistic nature, its success can be heralded by Bob eventual
measurement of it. The probability of the photon arriving at Bob's
side is $P_{0}=e^{-L/L_{\mathrm{att}}},$ where $L_{\mathrm{att}}=25$
km is the attenuation length of the channel with $c$ being that speed
of light in that channel. Now \textcolor{black}{after this interaction
with QM4} the photons degrees of freedom (polarization and timebin)
are swapped with each other (the $D_{\mathcal{L}}$ component is switched
with the $A_{\mathcal{S}}$ component). Then the photon interacts
with the last NV center (labelled QM2) resulting in the state (conditioned
on there being a photon at Bob's side): 

\noindent 
\begin{equation}
\begin{array}{c}
\left.|\phi_{12}^{+}\right\rangle \left.|\phi_{34}^{+}\right\rangle \left.|D_{\mathcal{S}}\right\rangle +\left.|\psi_{12}^{+}\right\rangle \left.|\phi_{34}^{+}\right\rangle \left.|A_{\mathcal{S}}\right\rangle \\
\;\;\;\;\;\;\;\;\;\left.+|\phi_{12}^{+}\right\rangle \left.|\psi_{34}^{+}\right\rangle \left.|D_{\mathcal{L}}\right\rangle +\left.|\psi_{12}^{+}\right\rangle \left.|\psi_{34}^{+}\right\rangle \left.|A_{\mathcal{L}}\right\rangle 
\end{array}\label{eq:4th_ent_step}
\end{equation}

\noindent where $\left.|\phi_{ij}^{+}\right\rangle =\left(\left.|g_{i}\right\rangle \left.|g_{j}\right\rangle +\left.|e_{i}\right\rangle \left.|e_{j}\right\rangle \right)$
and $\left.|\psi_{ij}^{+}\right\rangle =\left(\left.|g_{i}\right\rangle \left.|e_{j}\right\rangle +\left.|e_{i}\right\rangle \left.|g_{j}\right\rangle \right)$,
for $i=1(3),$ $j=2(4).$ Bob will then measure the photon (both polarization
and time bin's DOF) and so heralds its successful transmission. The
state of Eq. (\ref{eq:4th_ent_step}) will collapse in one of the
four tensor products of entangled states under ideal conditions. Depending
on the photons measurement result, bitflip operations can be performed
on Bob qubits ensuring that Alice and Bob share the required state
$\left.|\phi_{12}^{+}\right\rangle \left.|\phi_{34}^{+}\right\rangle $.
Appendix D shows the situation with channel loss with more detail.

\subsection{Advantages of the QMUXING entangling scheme}

\noindent The main advantage of the QMUXING entangling scheme is that
we only need one single photon to create two entangled pairs as compared
to the conventional schemes where at least two single photon sources
are needed. This means that we reduce the waiting time for entangling
both pairs. In fact, in the QMUXING scheme the time to entangle that
two pairs of memories is given by $2L/c,$ where $2L/c$ is simply
the time for Alice to send her photon to Bob and Bob to return a success/failure
message. In the case of success Alice and Bob can now use the entangled
pairs whereas in the case of failure, Bob needs to send a message
to Alice indicating another attempt is required including reinitializing
of the quantum memories. \textcolor{black}{The} reduced waiting time
for the QMUXING scheme means the quantum memories will dephase less
and so will result in higher fidelity pairs being generated. It is
straightforward to show that each pair of Fig. \ref{fig:QM_4qbits}(b)
will dephase simultaneously as 
\begin{equation}
\rho_{ij}^{\mathrm{dph}}(F_{ij})=F_{ij}|\phi_{ij}^{+}\left\rangle \negmedspace\right.\left.\negmedspace\right\langle \phi_{ij}^{+}|+(1-F_{ij})Z_{ij}|\phi_{ij}^{+}\left\rangle \negmedspace\right.\left.\negmedspace\right\langle \phi_{ij}^{+}|Z_{ij}\label{eq:rho_dep_QMUX}
\end{equation}
with $Z_{ij}$ being the $Z$ Pauli operator while $F_{ij}=F=\left(1+e^{-\frac{3L}{cT_{2}}}\right)/2$
is the fidelity of the generated entangled state. In the latter fidelity
expression, the term $3L/c$ takes into account the NV centers dephasing
time for the photon to traveling from Alice to Bob and the heralding
of a successful photon transmission to be communicated back to Alice
while $T_{2}$ is the coherence time of the memory. \textcolor{black}{For
the four QMUXING scheme}\textcolor{red}{{} }the state of the two entangled
pairs can be written as

\noindent 
\begin{equation}
\begin{array}{c}
\rho_{1234}^{\mathrm{dph}}(F)=\rho_{12}^{\mathrm{dph}}(F)\otimes\rho_{34}^{\mathrm{dph}}(F).\end{array}\label{eq:rho_QMUX_deph_4qubits}
\end{equation}

Now let us investigate the conventional entangling scheme \cite{Quantumcomm}
as it leads to a different dephasing process as the entangled states
are created at different times. Once the first entangled pair is created
one must wait until the second pair has been created before further
operations can be attempted. If we assume that the first(second) entangled
pair created is $\rho_{34},$ $\rho_{12,}$ respectively, then the
dephasing operation gives $\rho_{34}^{\mathrm{dph}}(F_{34}^{'}),$
$\rho_{12}^{\mathrm{dph}}(F_{12})$ with $F_{34}^{'}=\left(1+e^{-\frac{L}{cT_{2}}-\frac{2L}{cP_{0}T_{2}}}\right)/2,$
while $F_{12}$ is given by the expression above. The overall state
for both entangled pair is then $\begin{array}{c}
\rho_{1234}^{\mathrm{dph}}(F,F^{'})=\rho_{12}^{\mathrm{dph}}(F)\otimes\rho_{34}^{\mathrm{dph}}(F^{'})\end{array}$.

Next the QMUXING entangling scheme is not restricted to 2 entangled
pairs and can easily be extended to create $N$ entangled pairs of
NV centers separated by a distance (of course there are practical
limitations to this). In this case, a single photon will interact
with all Alice's qubits and then, after ideally travels across the
channel, will interact with Bob's qubits. In general, for creating
$N$ entangled pairs separated by a distance $L$, we need to encode
the photon into $N-1$ DOFs.\textcolor{black}{{} The photon components
will be coupled in such a way to create the final state given by the
sum of tensor} products of entangled pairs between Alice and Bob.

The advantages of using the QMUXING entangling scheme is further increased
for entangling a larger number of pairs, since the number of photons
is independent on the number of pairs, as in a conventional entangling
scheme. These pairs can then be used in further quantum tasks. 

\section{Application of quantum multiplexing to entanglement purification}

Let us now describe the method for generating high fidelity entangled
states. A\textcolor{blue}{{} }\textcolor{black}{general purification
protocol consists of performing local operations on $m$ entangled
qubits pairs in order to create $n<m$ pairs with a higher fidelity
than the initial pairs. One of the well known protocols to generate
high fidelity pairs is the Deutsch purification protocol \cite{QP2Deutsch}
where Alice and Bob share two copies of the Bell diagonal state}

\noindent 
\begin{equation}
\rho=A\rho_{\psi^{+}}+B\rho_{\psi^{-}}+C\rho_{\phi^{+}}+(1-A-B-C)\rho_{\phi^{+}},\label{eq:Bell_state}
\end{equation}
with $\rho_{\psi^{\pm},\phi^{\pm}}$ being the density matrices associated
with the Bell states, $|\psi^{\pm}\left\rangle =|10\left\rangle \right.\pm|01\left\rangle ,\right.\right.$
and $|\phi^{\pm}\left\rangle =|11\left\rangle \right.\pm|00\left\rangle \right.\right.$.
The coefficients $A,$ $B,$ and $C$ are constrained to give positive
real eigenvalues with $Tr\rho=1$. Alice and Bob begin their purification
protocol by applying an $X$ rotation on both their qubits, followed
by CNOT operations on both sides. Once these have been performed,
the target qubits are measured in the computational basis and the
result is communicated classically. If the protocol is successful,
the fidelity of the resulting state will be higher. In order to increase
even further the fidelity of the state, Alice and Bob can iterate
this procedure on two pairs having the same fidelity until they share
a high fidelity entangled state. 

\textcolor{black}{An alternative way of creating high fidelity entangled
pairs relies on error correction protocols. A conventional three qubit
error correction protocol works as follows. Alice and Bob create three
entangled pair and perform CNOT operations between the control qubits
and the target qubits. They measure the target qubits and communicate
classically the results of the measurements to each other. Depending
on the outcomes they apply a specific logic gate on their control
qubits to get the desired Bell state.}\textcolor{red}{{} }

\subsection{\label{subsec:Qmulti_3m}The QMUXING entangling scheme applied to
the Deutsch purification protocol}

Let is now apply the QMUXING entangling scheme to the Deutsch purification
protocol. We use the same procedure to create the entangled state
of Eq. (\ref{eq:4th_ent_step}) but then perform a Hadamard operation
on the qubits. In this case, upon a successful photon transmission,
our resulting state has the form 

\noindent 
\begin{equation}
\begin{array}{c}
\left.|\varphi_{12}^{+}\right\rangle \left.|\varphi_{34}^{+}\right\rangle \left.|D_{\mathcal{S}}\right\rangle +\left.|\chi_{12}^{+}\right\rangle \left.|\varphi_{34}^{+}\right\rangle \left.|A_{\mathcal{S}}\right\rangle \\
\;\;\;\;\;\;\;\;\;\left.+|\varphi_{12}^{+}\right\rangle \left.|\chi_{34}^{+}\right\rangle \left.|D_{\mathcal{L}}\right\rangle +\left.|\chi_{12}^{+}\right\rangle \left.|\chi_{34}^{+}\right\rangle \left.|A_{\mathcal{L}}\right\rangle 
\end{array}\label{eq:Ent_state_pur}
\end{equation}
with $\left.|\varphi_{ij}^{+}\right\rangle =\left(\left.|+_{i}\right\rangle \left.|+_{j}\right\rangle +\left.|-_{i}\right\rangle \left.|-_{j}\right\rangle \right)$
and $\left.|\chi_{ij}^{+}\right\rangle =\left(\left.|+_{i}\right\rangle \left.|-_{j}\right\rangle +\left.|-_{i}\right\rangle \left.|+_{j}\right\rangle \right)$,
where $(i,\,j)=(1,\,2),$ $(3,\,4),$ respectively.\textcolor{black}{{}
Bob will measure the photon and will flip his qubits depending on
the photon outcome as described in Sec.} \ref{sec:quantum_multi_scheme}\textcolor{black}{.
Alice and Bob then} will perform a CNOT operation between their respective
qubits and will measure the target qubits. They will communicate the
outcomes of the target qubits to each other and they will keep the
entangled pair if the outcomes are the same, otherwise they will start
the protocol again.
\begin{figure}
\begin{centering}
\includegraphics[scale=0.14]{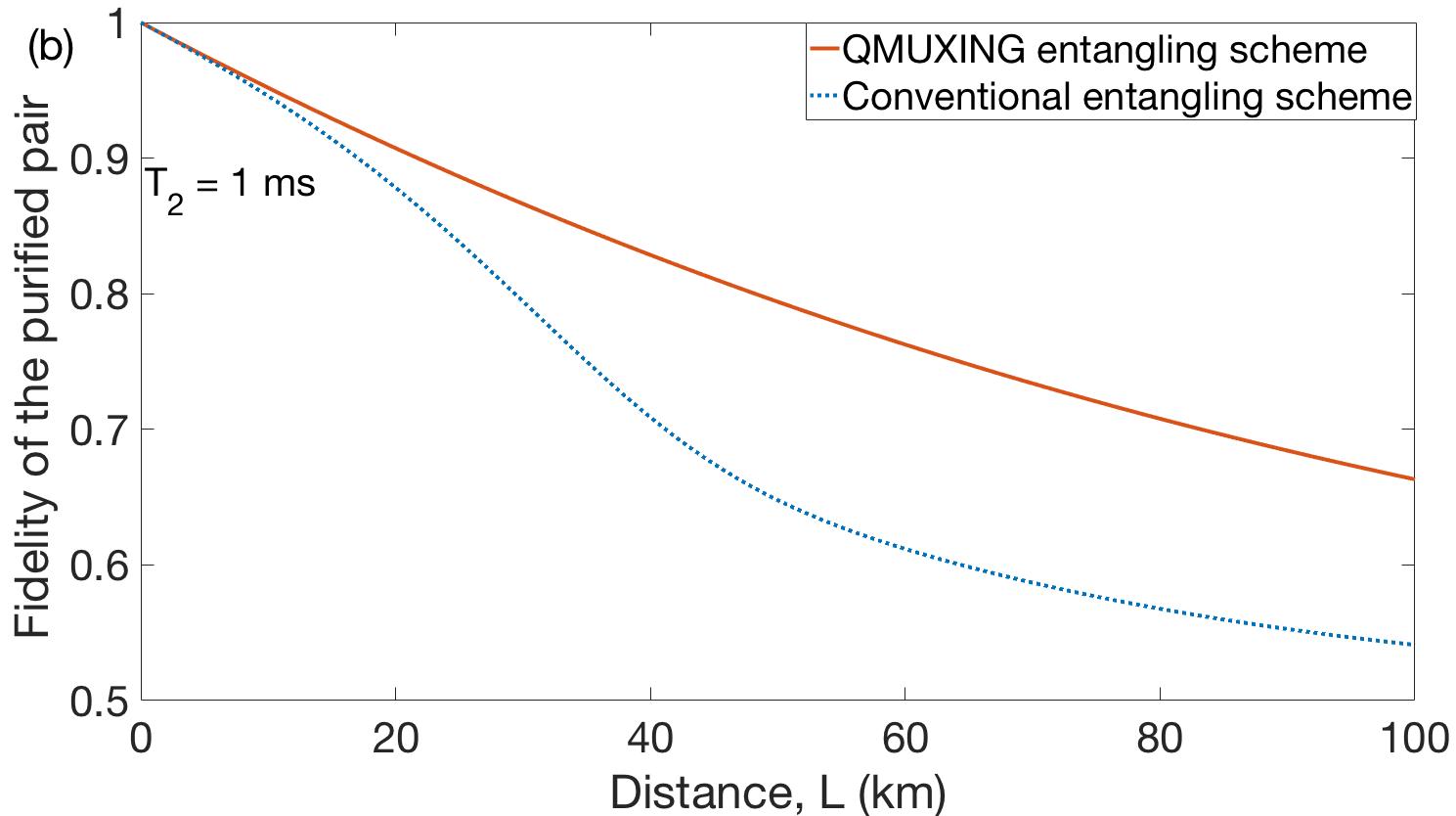}
\par\end{centering}
\caption{\label{fig:Fidelities-1}In (a), entanglement purification scheme
of two entangled pairs and in (b) fidelities of the purified pair
obtained through the Deutsch protocol in which the entangled pairs
have been created with the QMUXING entangling scheme (solid lines)
and with a traditional prepare and measure entangling scheme (dotted
line) for a coherence time, $T_{2}=1$ ms. }
\end{figure}

Compared to the Deutsch purification scheme in which the quantum memories
are created in the traditional way, our approach is faster due to
the fact that both the entanglement creation and purification are
acknowledged simultaneously. While the photon is transmitted through
the channel both pairs will go through a dephasing quantum channel
given by $\rho_{0}^{\mathrm{deph}}(F_{0})$ (see Eq. \ref{eq:rho_dep_QMUX})
where $F_{0}=$$\left(1+e^{-L/cT_{2}}\right)/2$. At this point a
CNOT gate is applied and the target qubits are measured. The fidelity,
$F_{\mathrm{QMX}},$ of the purified pair will be given by 

\noindent 
\begin{equation}
F_{\mathrm{QMX}}=\frac{F_{0}^{2}}{F_{0}^{2}+(1-F_{0})^{2}}.
\end{equation}
Now, Alice and Bob will communicate classically the outcomes of their
measurement therefore their state, $\rho_{\mathrm{QMX}}^{\mathrm{dph}}$,
will dephase to $\rho_{\mathrm{QMX}}^{\mathrm{dph}}(F_{\mathrm{QMX}}^{\mathrm{dph}}),$
where $F_{\mathrm{QMX}}^{\mathrm{dph}}=\left(1+(2F_{\mathrm{QMX}}-1)e^{-2L/cT_{2}}\right)/2$
(see Appendix E for details). 

Now in the traditional Deutsch protocol where the entanglement distribution
is done independently per pair the fidelity of the purified pair will
be given by 

\noindent 
\begin{equation}
F_{\mathrm{trad}}=\frac{F_{12}F_{34}^{'}}{F_{12}F_{34}^{'}+(1-F_{12})(1-F_{34}^{'})},
\end{equation}
where $F_{12}$ and $F_{34}^{'}$ have been defined in Sec. II. After
the entangled pairs are created, Alice and Bob will perform the local
operations and will measure their target qubits. Then they will communicate
the outcomes of the measurement to each other. During this time, the
purified pair state, $\rho_{\mathrm{trad}}^{\mathrm{dph}},$ will
dephase to $\rho_{\mathrm{trad}}^{\mathrm{dph}}(F_{\mathrm{trad}}^{\mathrm{dph}}),$
where $F_{\mathrm{trad}}^{\mathrm{dph}}=\left(1+(2F_{\mathrm{trad}}-1)e^{-2L/cT_{2}}\right)/2.$

Now in Figure \ref{fig:Fidelities-1}(b) we compare $F_{\mathrm{QMX}}^{\mathrm{dph}}$
and $F_{\mathrm{trad}}^{\mathrm{dph}}$ versus $L$ where we have
set a $T_{2}=1$ ms coherence time for the memories. We observe that
for short distances the two fidelities are almost the same. However,
at distances larger than $20$ km the fidelity of the entangled pair
created by the QMUXING scheme is much higher compared to the one of
the entangled pair created with the traditional entangling scheme.
This advantage is biggest at $L=50$ km where $F_{\mathrm{QMX}}^{\mathrm{dph}}=0.8$
and $F_{\mathrm{trad}}=0.6.$

\subsection{The QMUXING protocol}

We illustrate here a method, which has been introduced in \cite{private},
to create high fidelity entangled pairs through the QMUXING entangling
scheme with a built-in purification protocol. We call this method
QMUXING protocol. A remarkable feature of encoding a photon into multiple
DOFs is that the DOFs correspond to qubits on which we might perform
the same local operations applied above. This in turn means the number
of matter qubits can be reduced. 
\begin{figure}
\begin{centering}
\includegraphics[scale=0.31]{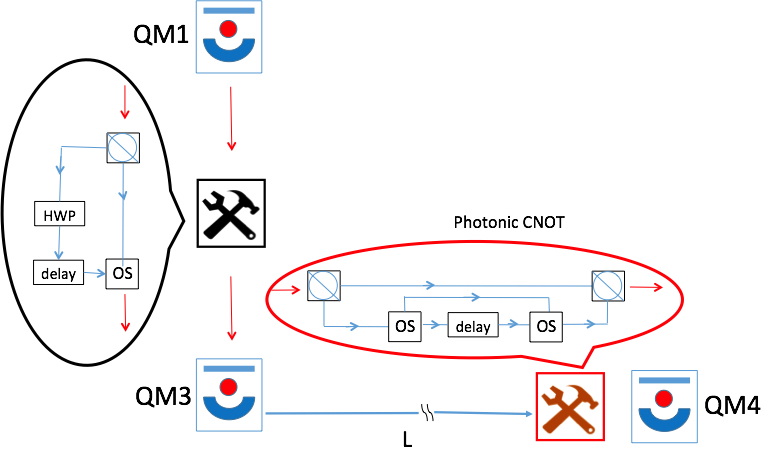}
\par\end{centering}
\caption{\label{fig:QMUX3qubits}Three qubit quantum multiplexing protocol.
The first two steps are identical as the ones described in Fig. 1.
The spin states of the NV centers are rotated into the diagonal basis.
When the photon being transmitted across the channel arrives at Bob
a CNOT operation is performed between QM3 and QM1 and between the
photonic modes of the photon through the ``photonic CNOT'' operation
of the Figure. The photon then interacts with the last NV center and
it is measured. The protocol is successful if Alice measures ``$+(-)"$
and Bob will measures ``$\mathcal{S}(\mathcal{L})"$ on the photons
degree of freedom. In any other case, the protocol is aborted and
Alice and Bob start again.}

\end{figure}

In this protocol, the state of our system, after the photon has been
interacted with the memories QM1 and QM3 of Fig. \ref{fig:QMUX3qubits},
is given by Eq. (\ref{eq:3rd_ent_step}). A Hadamard applied to Alice's
qubit gives:

\noindent 
\begin{equation}
|++D_{\mathcal{S}}\left\rangle +|+-A_{\mathcal{S}}\left\rangle +\right.|-+D_{\mathcal{L}}\left\rangle +\right.|--A_{\mathcal{L}}\left\rangle .\right.\right.
\end{equation}

\noindent We can now apply a CNOT operation between QM3 (control)
and QM1 (target) and between the polarization DOF (control) and the
TB DOF (target). This CNOT operation works as follows: the diagonal
component will leave unaffected the time-bin component and the antidiagonal
component will flip the time-bin component. The CNOT on the photonic
qubits can be implemented by the scheme represented in the the ``photonic
CNOT'' operation of Fig. \ref{fig:QMUX3qubits}. Upon a successful
transmission of the photon through the channel (which will be heralded
by the photon measurement), the photon then interacts with QM4, which
is successively rotated in the diagonal basis. The final resulting
state has the form:

\noindent 
\begin{equation}
\begin{array}{c}
\begin{array}{c}
|+_{1}\left\rangle \negmedspace\right.|\varphi_{34}^{+}\left\rangle \negmedspace\right.|D_{\mathcal{S}}\left\rangle \negmedspace\right.+|+_{1}\left\rangle \negmedspace\right.|\chi_{34}^{+}\left\rangle \negmedspace\right.|A_{\mathcal{S}}\left\rangle \negmedspace\right.\end{array}\\
\;\;\;\;\;\;\;\;\;+|-_{1}\left\rangle \negmedspace\right.|\chi_{34}^{+}\left\rangle \negmedspace\right.|A_{\mathcal{L}}\left\rangle \negmedspace\right.+|-_{1}\left\rangle \negmedspace\right.|\varphi_{34}^{+}\left\rangle \negmedspace\right.|D_{\mathcal{L}}\left\rangle \negmedspace\right.
\end{array}
\end{equation}

\noindent Alice with now measure QM1 in the diagonal basis while at
the same time Bob measures the state of the photon (both degrees of
freedom). They communicate the results with each other via the classical
communications channel. A purified pair is obtained if the outcomes
of QM1 is ``$+(-)"$ and of the TB DOF is ``$\mathcal{S}(\mathcal{L})",$
respectively. In this case the states will be given by $|\varphi_{34}^{+}\left\rangle ,\right.$
$|\chi_{34}^{+}\left\rangle \right.$ respectively. For the ``$+\mathcal{L}"$
and ``$-\mathcal{S}"$ results the protocol has failed and we need
to begin again with the entanglement distribution. Of course this
considerations have not included dephasing yet. It can be simply handled
(appendix F) and for instance with the ($+$, $\mathcal{S}$) measurement
result, our quantum state would have the form

\noindent \textcolor{black}{\footnotesize{}
\begin{equation}
\begin{array}{c}
\negthickspace\negthickspace\negthickspace\rho_{\mathrm{3}}=\frac{F^{2}}{F^{2}+(1-F)^{2}}|\varphi_{34}^{+}\left\rangle \negmedspace\right.\left.\negmedspace\right\langle \varphi_{34}^{+}|+\frac{(1-F)^{2}}{F^{2}+(1-F)^{2}}X_{3}|\varphi_{34}^{+}\left\rangle \negmedspace\right.\left.\negmedspace\right\langle \varphi_{34}^{+}|X_{3},\end{array}
\end{equation}
}where $X_{3}$ is the $X$ Pauli operator applied to QM3. 

We can also generalize the QMUXING protocol to a larger number of
memories. In this case, if $N$ is the total number of pairs of a
conventional protocol, the total number of matter qubits that we need
QMUXING protocol will be given by $N+1,$ since we need $N-1$ effective
extra DOFs are needed to entangled $N$ pairs (this can be also ex
TB modes). 

We can further reduce the number of matter qubits if we use the nuclear
spin of an NV center as a qubit. We can in fact transfer the electron
spin state of the NV center into the nuclear spin after the first
interaction of the photon. In this way, the photon can interact again
with the electron spin and then travel across the channel where it
will interact with Bob's qubit (see Appendix G).

\section{Performance metrics}

It is important now to investigate quantitatively what improvements
this new scheme gives. As the main figure of merit, we will calculate
the rate at which Alice and Bob can share purified entangled states
normalized by the number of physical resources (the number of matter
qubits or quantum memories and the average number of single photons
needed to create the entangled states). In order to evaluate the impact
of the number of resources used, we introduce the cost functions $C_{M}$
and $C_{p},$ which multiply the number of matter qubits and the average
number of single photons, respectively. 

\subsection{Normalized purification rates}

For a purification protocol with $k$ purification rounds our normalized
rate is defined by:

\noindent 
\begin{equation}
R_{\alpha}(k)=\frac{r_{\alpha}(k)}{M_{\alpha}(k)C_{M}+m_{\alpha}(k)C_{p}},\label{eq:rate}
\end{equation}
where $r_{\alpha}(k)$ is the raw rate for establishing a high fidelity
entangled pair over a distance $L$ and with the subscript $\alpha=\mathrm{QMX}$
($\alpha=\mathrm{D}$) referring to the QMUXING protocol (Deutsch
protocol with traditional entanglement creation), respectively. The
rate of establishing purified pairs after $k$ purification rounds
using the QMUXING protocol is given by $R_{\mathrm{QMX}}(k)$ with

\noindent 
\begin{equation}
\left(r_{\mathrm{QMX}}(k)\right)^{-1}=\frac{2}{c}\frac{L}{P_{0}\prod_{I=1}^{k}P_{D}(i)},\label{eq:our_rate}
\end{equation}

\noindent where $P_{D}$ is the probability of a successful purification
round with $M_{\mathrm{QMX}}(k)=2^{k}+1$ and $m_{\mathrm{QMX}}(k)=1/P_{0}.$
The rate $r_{\mathrm{QMX}}(k)$ is given by only one term associated
with the time the photon travels on the optical fiber and reach Bob
side and the time for classical communication. The number of matter
qubits is less than the traditional purification scheme due to the
local operations performed on the extra DOFs of the photon. Similarly
the rate, $R_{\mathrm{D}},$ of the Deutsch protocol for a fixed number,
$k,$ of distillation rounds which entangle pairs generated conventionally
is given by $R_{D}(k)$ with

\noindent 
\begin{equation}
\begin{array}{c}
\left(r_{\mathrm{D}}(k)\right)^{-1}=\left(\frac{3}{2}\right)^{k}\frac{2}{c}\frac{L}{P_{0}\prod_{I=1}^{k}P_{D}(i)}\\
+\left(\frac{3}{2}\right)^{k-1}\frac{L}{c\prod_{I=1}^{k}P_{D}(i)}+...\frac{L}{cP_{D}(1)},
\end{array}\label{eq:Deutsch_rate}
\end{equation}
where $M_{\mathrm{D}}(k)=2^{k+1}$ and $m_{\mathrm{D}}(k)=2^{k}/P_{0}$.
The first term is the rate to establish an entangled state over a
distance $L$ and to communicate classically that the entangling step
has been successful. The factor $3/2,$ is the average waiting time
to prepare two entangled pairs. The other terms are associated with
the times to acknowledge that the $k-$th distillation rounds has
occurred. These latter terms are not present in the rate of our QMUXING
protocol as the purification steps are performed during the time to
establish the entanglement. 

Next to quantify the improvement we have, let us calculate the ratio
of the rates, $R_{\mathrm{QMX}}/R_{\mathrm{D}},$ for $k=1$ and $k=2,$
respectively. In order to analyze the effect of the number of the
physical resources, we vary the weighting coefficients $\frac{}{}C_{M}$
and $C_{p}.$ 
\begin{figure}
\begin{centering}
\includegraphics[scale=0.15]{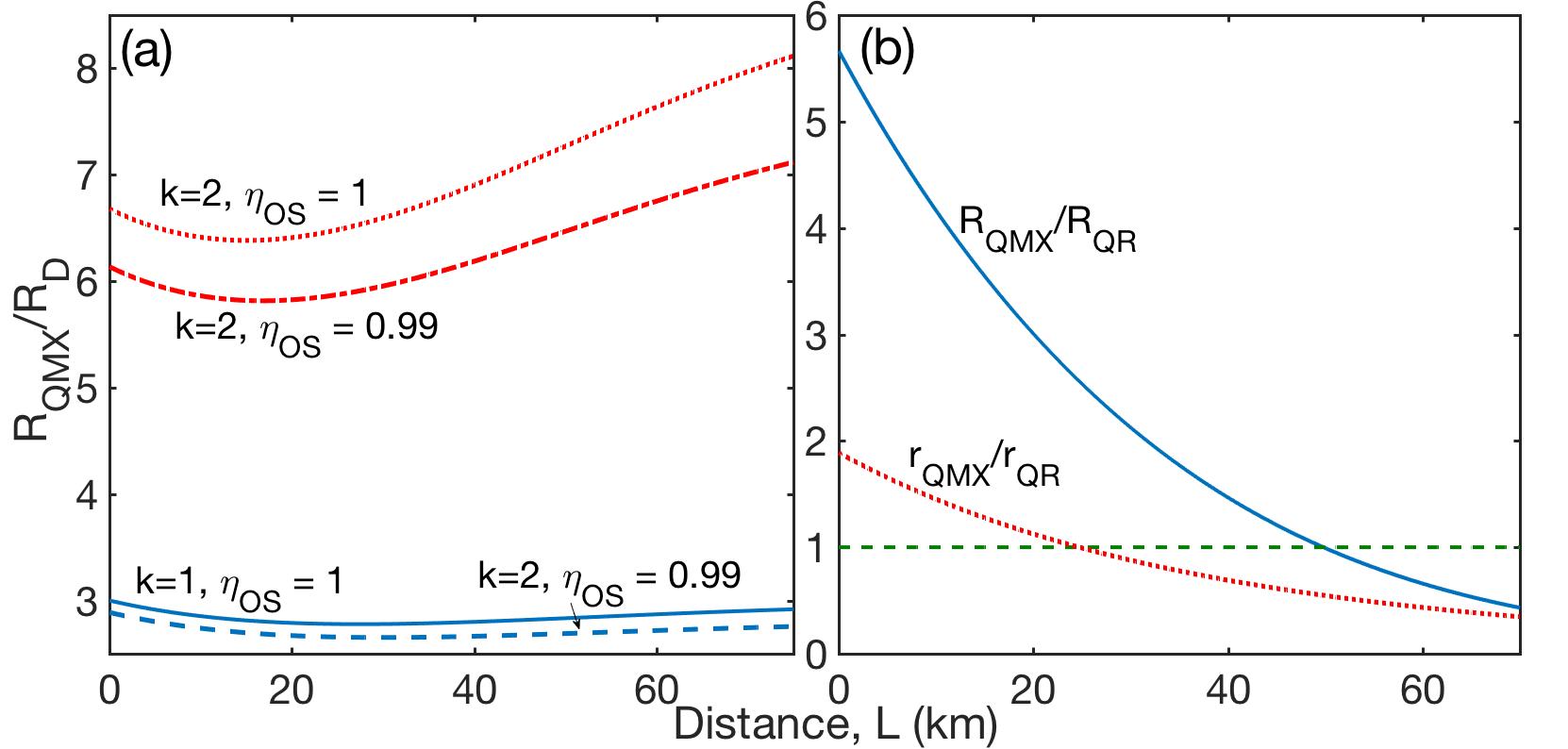}
\par\end{centering}
\caption{\label{fig:ratio_no_err}Ratio (a) between the normalized rates of
the QMUXING protocol and the Deutsch protocol for one and two distillation
rounds with perfect (dotted line and solid line, respectively) and
imperfect (dash-dotted line and dashed line, respectively) optical
switches, respectively and (b) between the three qubit QMUXING protocol
and a conventional single node quantum repeater scheme. In (b), the
intersections of the curves with the green dashed line show the crossover
distances. We modeled the imperfection of the optical switches as
a loss even with transmission probability $\eta_{OS}=0.99.$}
\end{figure}
 In Fig. \ref{fig:ratio_no_err}(a) we plot $R_{\mathrm{QMX}}/R_{\mathrm{D}}$
versus the distance for $k=1$ and $k=2$ in the ideal case of perfect
optical switches. The two curves show an average increment of the
rate equal to $2.5$ and 7 for a total distance of $L=70$ km, respectively,
compared to $R_{D}$. They both have a minimum value at $L=25$ km
and $L=15$ km, respectively. In fact, if we consider the rates for
the case of $k=1,$ for $C_{M}=C_{p}=1,$ this ratio can be expressed
as:

\noindent 
\begin{equation}
\frac{R_{\mathrm{QMX}}}{R_{\mathrm{D}}}=\frac{r_{\mathrm{QMX}}}{r_{\mathrm{D}}}\frac{4P_{0}+2}{3P_{0}+1}=\left(\frac{3}{2}+\frac{P_{0}}{2}\right)\frac{4P_{0}+2}{3P_{0}+1}.\label{eq:ratio_k1}
\end{equation}
This shows that the improvement of our protocol is then given by two
factors: the increment of the rate of establishing an entangled state,
which depend on the factor $3/2$ and on the distance between the
users. Additionally, the latter term is multiplied by the ratio between
the number of resources, which are less in our protocol. The ratio
between the raw rates decreases exponentially with the distance as
well as the ratio between the number of resources increases exponentially
with the distance. Therefore, we expect a minimum increment in the
ratio at certain distance. The $k=2$ case follows a similar explanation.
Figure \ref{fig:ratio_no_err}(a) shows also the case in which the
optical switches efficiency is 0.99. As expected for $k=2$ the difference
between the two curves is higher than the one for $k=1$ due to the
higher number of optical switches needed.

We can also compare the rate of the three-qubit QMUXING protocol with
the rate of a single node quantum repeater protocol in which the pairs
has been purified at a distance $L_{0}=L/2.$ For such a system the
rate of creating a purified pair over a distance $L$ is given by
$R_{\mathrm{QR}}=r_{\mathrm{QR}}/(8C_{M}+4C_{p}/P_{0}^{1/2}$) with

\noindent {\footnotesize{}
\begin{equation}
(r_{\mathrm{QR}})^{-1}=\left(\frac{3}{2}\right)^{2}\frac{2}{c}\frac{L/2}{P_{0}^{1/2}P_{D}P_{ES}}+\frac{3}{2}\frac{L/2}{cP_{D}P_{ES}}+\frac{L/2}{cP_{ES}},
\end{equation}
}{\footnotesize \par}

\noindent where $P_{ES}$ is the probability of a successful entangling
swapping operation (assumed to be 0.9 here). Figure \ref{fig:ratio_no_err}(b)
shows the ratio of both the normalized rate and the raw rate of the
QMUXING protocol against the standard one node quantum repeater protocol,
in which the pairs have been purified over a distance $L/2.$ The
normalized(raw) rate of the QMUXING protocol outperforms the one of
the single node QR up to a distance $L\sim50$(25) km (see Fig. \ref{fig:ratio_no_err}(b).
We can also apply the QMUXING protocol to a single node QR scheme.
In this case, our rate outperforms the conventional one for all distances.

\subsection{Normalized rate in the error correction scheme}

\noindent For an $N$-qubits error correction protocol the normalized
rate is similarly given by: 

\noindent 
\begin{equation}
R_{\alpha}^{\mathrm{EC}}(N)=\frac{r_{\alpha}^{\mathrm{EC}}(N)}{M_{\alpha}(N)C_{M}+m_{\alpha}(N)C_{p}},
\end{equation}

\noindent where $r_{\alpha}^{\mathrm{EC}}(N)$ is the raw rate for
creating a high fidelity pair. The rate of the three-qubit error correction
QMUXING protocol is given by $R_{\mathrm{QMX}}^{\mathrm{EC}}$ with
$\begin{array}{c}
\left(r_{Q\mathrm{MX}}^{E\mathrm{C}}\right)^{-1}=\frac{2}{c}\frac{L}{P_{0}},\end{array}$ $M_{\mathrm{QMX}}(3)=4$ and $m_{\mathrm{QMX}}(3)=1/P_{0}.$ The
rate of the three-qubit error correction protocol in which the pairs
are created in the conventional way is given by $R_{\mathrm{D}}^{\mathrm{EC}}$
with $r_{\mathrm{D}}^{\mathrm{EC}}=1.7\frac{2}{c}\frac{L}{P_{0}}$
and $m_{\mathrm{D}}(3)=3/P_{0}$. The factor $1.7/P_{0}$ is an approximative
value for the average time we have to wait in order entangle three
pairs (Appendix \ref{sec:P0approx}).

\begin{figure}
\begin{centering}
\includegraphics[scale=0.16]{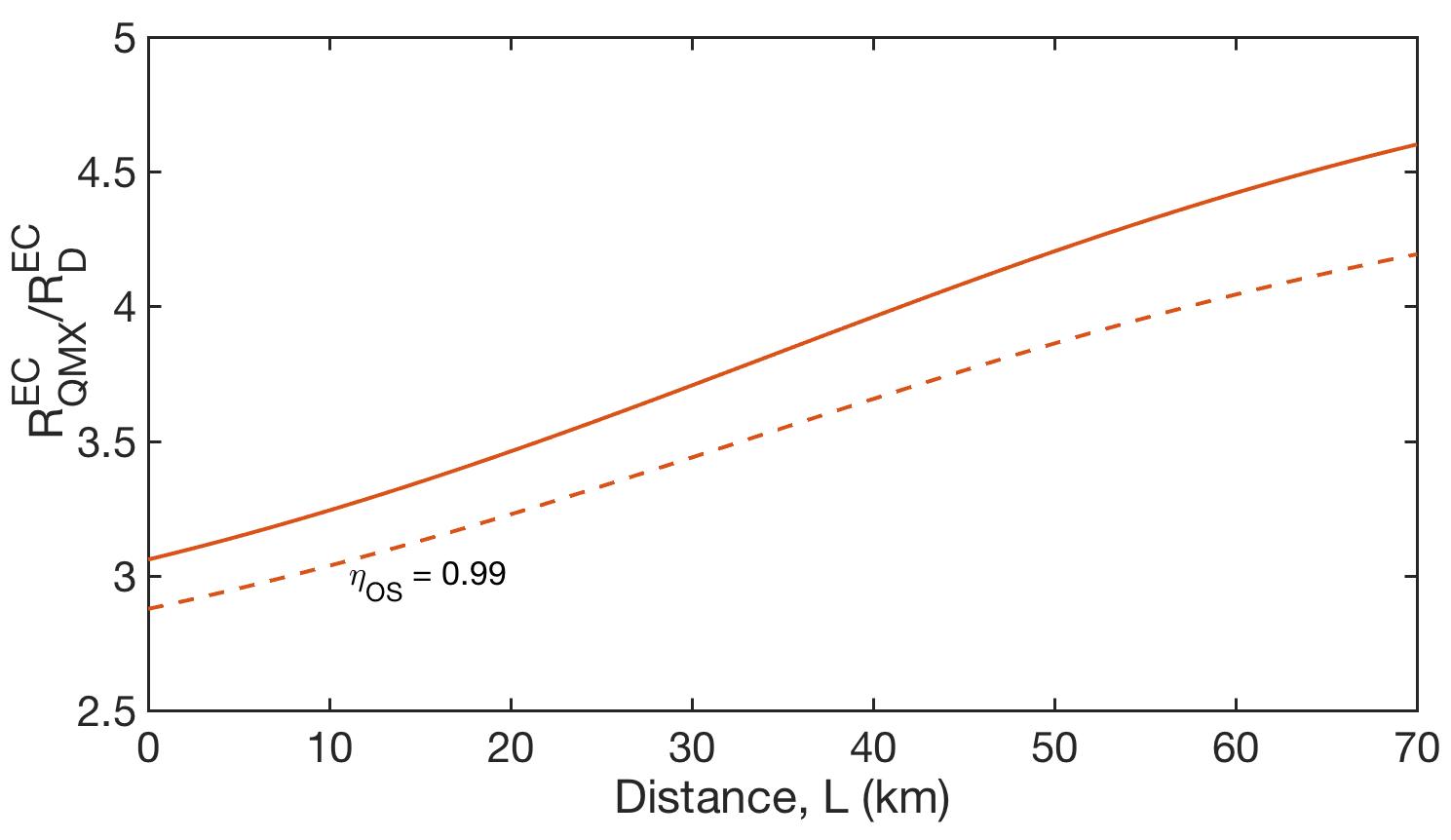}
\par\end{centering}
\caption{\label{fig:ratioEC}Ratio between the rate of the 3-qubit QMUXING
protocol with built-in error correction protocol and the rate of the
3-qubit error correction protocol in which the pairs are created in
the conventional way perfect (solid line) and imperfect optical switches
(dashed line).}
\end{figure}

In Fig. \ref{fig:ratioEC} we plot $R_{\mathrm{QMX}}^{\mathrm{EC}}/R_{\mathrm{D}}^{\mathrm{EC}}$.
In this case, the ratio between the raw rates is constant as shown
in Fig. \ref{fig:ratioEC} and it increases with the distance reaching
a value of $5$ at $L=70$ km. By considering imperfect optical switches
this ration is a bit lower as shown in the dashed line of Fig. \ref{fig:ratioEC}.

\begin{figure}
\begin{centering}
\includegraphics[scale=0.14]{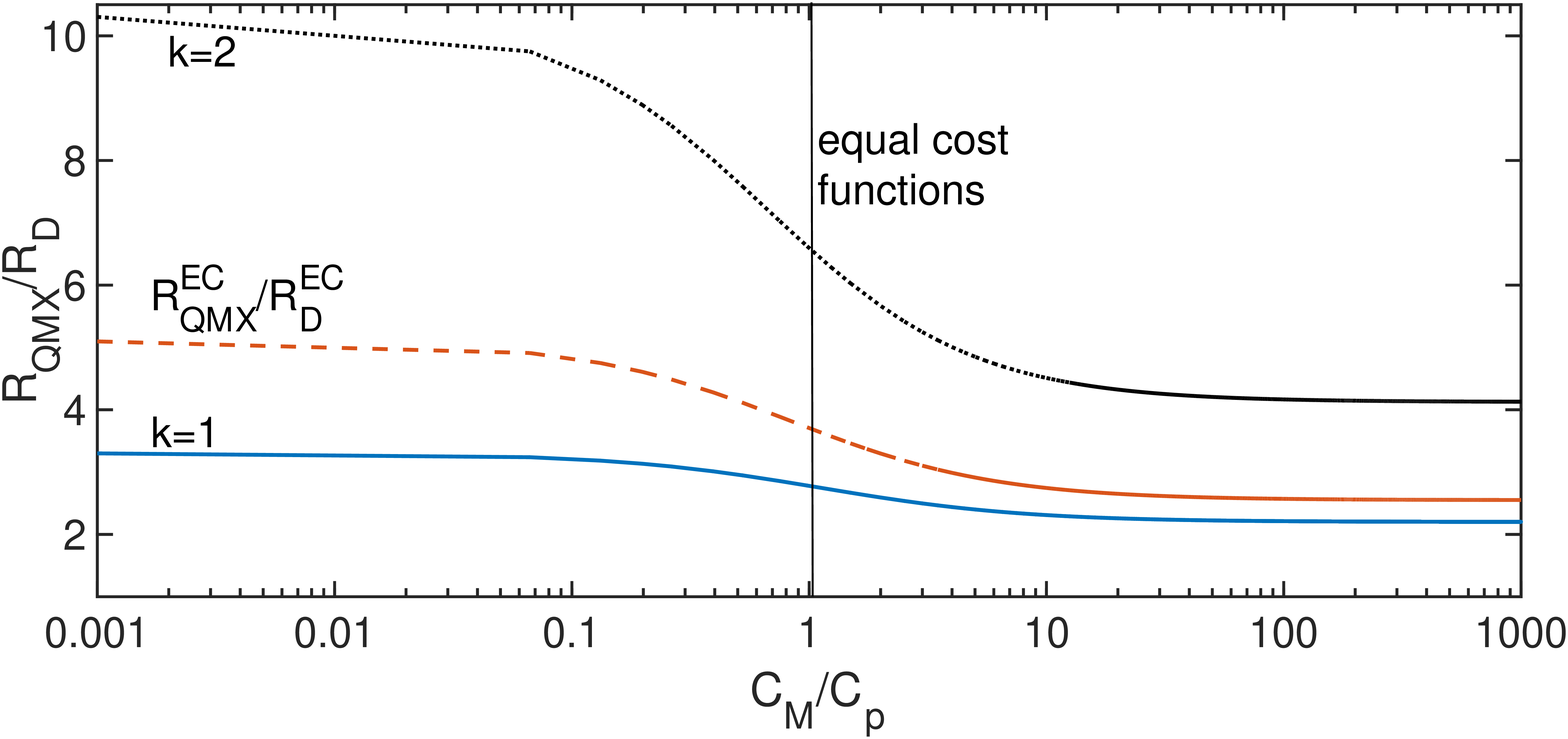}
\par\end{centering}
\caption{\label{fig:ratio_vs_C}Ratio between the cost functions $C_{M}$ and
$C_{p}$ at a distance $L=30$ km. The vertical line corresponds to
the case when $C_{M}=C_{p}.$ At a fixed distance between Alice and
Bob, higher (lower) values of $C_{p}$ respect to $C_{M}$ correspond
to a bigger(smaller) increment of the ratio of the two rates. }
\end{figure}

Since it is not straightforward to estimate the actual values of $C_{M}$
and $C_{p},$ we calculate the ratio of the normalized rates at a
fixed distance versus the ratio of the cost functions $C_{M}$ and
$C_{p},$ as illustrated in Fig. (\ref{fig:ratio_vs_C}). The point
$C_{M}=C_{p}=1,$ which correspond to the case of Fig. (\ref{fig:ratio_no_err}),
splits the $x-$axis into two parts. For $C_{M}<C_{p}$ the normalized
rate of our protocol achieves higher values compared to the case of
equal cost function. That means that when we include the number of
physical resources in the purification protocol, it is more convenient
using a less number of matter qubits than the average number of single
photons. The cost functions $C_{M}$ and $C_{p}$ might depends on
several factors, such as its effective commercial costs and other
characteristics. 

\section{Discussion and Conclusion}

In this work, we have introduced a new entangling scheme that allows
to create multiple entangled pairs by using only a single photon.
The photon, encoded in multiple degrees of freedom (DOFs), will entangle
a series of NV centers prepared locally through the scheme \cite{DE}.
By switching among the various degrees of freedom entangled pairs
between two remote users can be created. We call this new entangling
method ``quantum multiplexing'' (QMUXING), since the photon carries
multiple DOFs. The advantage of using such a method is the less number
of resources needed and the lower average waiting time for creating
an entangled pair. This will reduce the detrimental effects of the
decoherence effect on the quantum memories in use. 

We have also applied the quantum multiplexing method to the Deutsch
purification protocol and shown that the raw generation rate is faster
in our case compared to the conventional entangling scheme. We also
use the QMUXING entangling scheme to generate purified entangled pairs
with a built-in purification protocol on which the extra qubits of
the photon are used as effective qubits. In this way, for a given
number of purification rounds, $k,$ of a purification protocol in
which the entangled pairs have been created with a conventional entangling
scheme, we reduce the number of matter qubits when the QMUXING protocol
is in use. In order to estimate the rate at which Alice and Bob can
share a purified entangled pair after $k$ distillation rounds, we
use a normalized figure of merit that takes into account the raw entanglement
rate over the total number of physical resources, in terms of matter
qubits and average number of single photons. To each of these parameters
we associate a cost function in order to assess the impact of such
resources on the rate. We plot the ratio of the normalized rate of
our new protocol and the rate of creating high fidelity pairs through
the Deutsch protocol for $k=1$ and $k=2$ and for the three-qubit
error correction protocol when the entangled pairs are created with
a traditional method and when perfect optical switches are in use.
Initially, we consider that the cost functions are equal. We obtain
that our protocol is roughly 2.5 faster than the other purification
system for $k=1$ and up to $7$ times faster for $k=2.$ The QMUXING
with error correction built-in scheme is 4.5 times faster than the
conventional error correction protocol. For such a system, we calculate
the average waiting time of entangling three pairs and we extend this
calculation also for a number of pairs until $10.$ These values of
the average time can be used in a further work in order to estimate
the difference between the rate of a purification protocol having
$N$ entangled pairs and an error correction protocol with the same
number of qubits. We then calculate the ratio of the rate of the QMUXING
protocol and the rate of Deutsch protocol with pairs created with
a traditional scheme versus the ratio of the cost functions related
to the matter qubits, $C_{M},$ and to the average number of photons,
$C_{p},$ at a fixed distance. We find that our protocol shows a bigger
improvement when the photon cost is more expensive than the memories. 

Our QMUXING entangling scheme can also be applied to other quantum
communication protocols, such as a the multiple memories configuration
in a quantum repeater protocol, and to any protocol that requires
entanglement distribution between two remote users. Moreover, the
lower number of physical resources needed can dramatically reduce
the costs required for its implementation. 
\begin{acknowledgments}
NLP acknowledges support from the JSPS international fellowship. This
project was made possible through the support of a grant from the
John Templeton Foundation. The opinions expressed in this publication
are those of the authors and do not necessarily reflect the views
of the John Templeton Foundation (JTF \#60478). KN also acknowledges
support from the MEXT KAKENHI Grant-in-Aid for Scientific Research
on Innovative Areas \textquotedblleft Science of Hybrid Quantum Systems\textquotedblright{}
Grant No. 15H05870. 
\end{acknowledgments}

\appendix

\section{\label{sec:P0approx}Entanglement distribution time for multiple
pairs}

\begin{figure}
\begin{centering}
\includegraphics[scale=0.14]{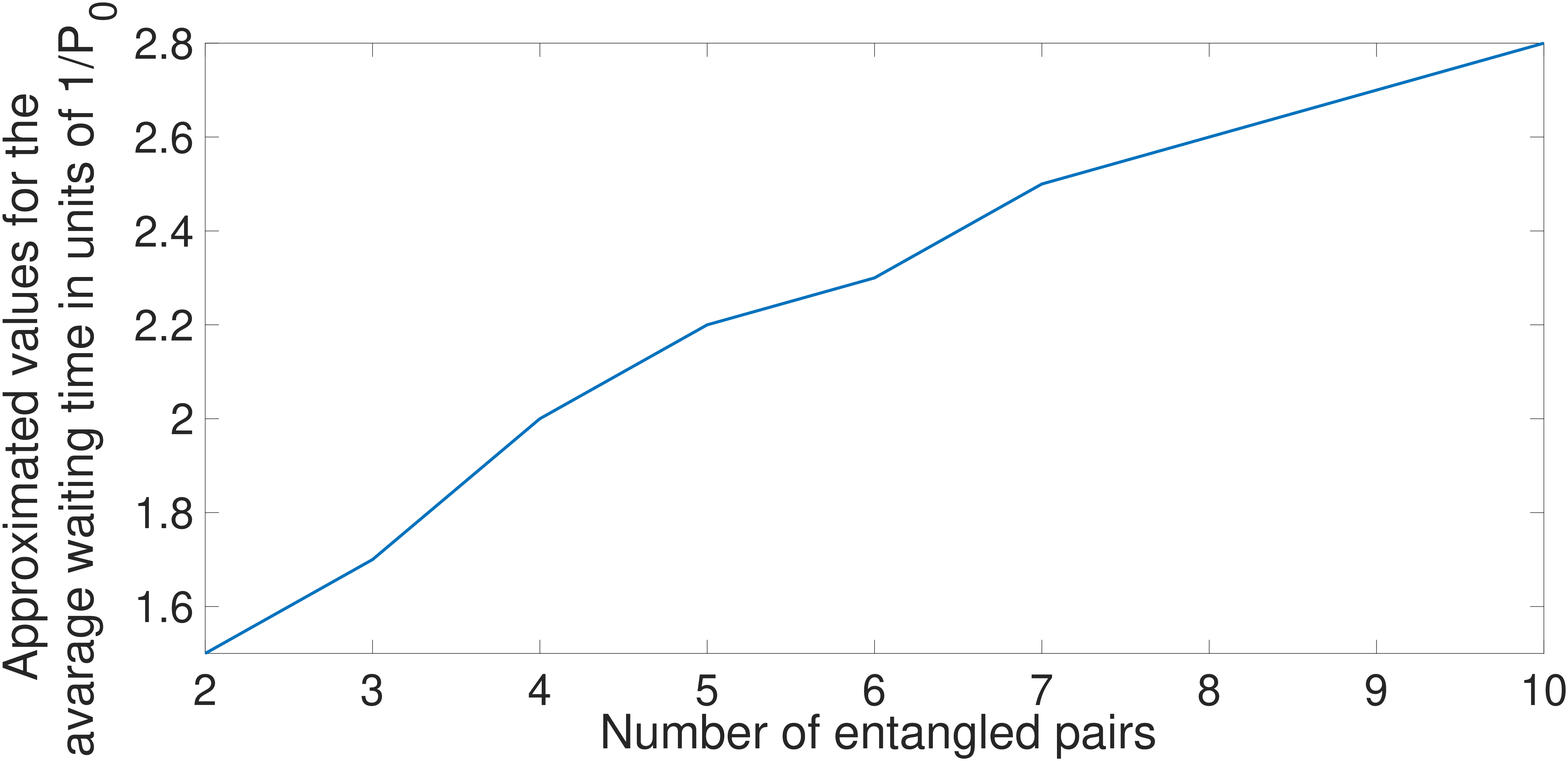}
\par\end{centering}
\caption{\label{fig:Approximative_p0}Approximative values in units of $1/P_{0}$
of the average waiting time for entangling $N=2$ to $N=10$ pairs
separated by a distance $L$. }

\end{figure}
Let us now derive the average waiting probability of establishing
three entangled pairs. We used this value for estimate the rate of
a the three-qubit error correction protocol described in Sec. IVB.

Given a success entanglement probability $P_{0},$ the distribution
probability of the number of attempts, $n,$ before we can establish
an entangled pair over an elementary link is given by \cite{P0}:

\noindent 
\begin{equation}
p(n)=(1-P_{0})^{n-1}P_{0}.
\end{equation}
The distribution probability for entangling three elementary links
will be given by:

\noindent 
\begin{equation}
\negthickspace q(n)=p(n)^{3}+3p(n)^{2}\sum_{k=1}^{n-1}p(k)+3p(n)\sum_{r=1}^{n-1}p(k)\sum_{s=1}^{n-1}p(k).
\end{equation}
whose expectation value is given by:

\noindent 
\begin{equation}
\left\langle n\right\rangle =\sum_{n=0}^{\infty}nq(n)=\frac{3P_{0}^{3}-12P_{0}^{2}+19P_{0}-11}{P_{0}(P_{0}^{2}-3P_{0}+3)(P_{0}-2)}.\label{eq:1.7factor}
\end{equation}
For small $P_{0},$ the expression in Eq. \ref{eq:1.7factor} can
be approximated by $\left\langle n\right\rangle \sim1.7/P_{0}$. In
addition to that, in the prepare and measure entangling scheme, the
average waiting time for entangling $N$ pairs will increase, as shown
in Fig. \ref{fig:Approximative_p0} for $N=3..10.$ By applying the
QMUXING entangling scheme on the other hand, if we neglect the interaction
time of the photon with the quantum memory, the average waiting time
for entangling $N$ pairs will still be proportional to $L/cP_{0}$
(we have not included the communication time). 

\section{Optical switch error model }

\textcolor{black}{In a real implementation of the QMUXING entangling
scheme, we have to consider the errors associated with the optical
switches, since they are the only new optical elements which are not
present in the conventional entangling scheme. We model this error
as a loss of the photon, with transmission coefficient given by $\eta_{OS}.$
We both consider the ideal case, in which $\eta_{OS}=1,$ and the
more realistic case, in which }$\eta_{OS}=0.99$. The number of times
the optical switches are used in the quantum multiplexing scheme depends
on the number of pairs $N$ we want to entangle. In particular, for
each pair creation we have to add a time-bin component by using the
gates of the black bubble of Fig. \ref{fig:QM_4qbits} and then, after
the photon has travelled across the channel, we have to perform the
operations of the green bubble of Fig. \ref{fig:QM_4qbits} for each
pair we want to entangle. Therefore, for $N$ entangled pairs we want
to create, the number of optical switches is given by $\frac{3}{2}N-3.$
Therefore, we substitute $P_{0}$ with $P_{0}^{'}=\eta_{OS}^{\frac{3}{2}N-3}P_{0}.$ 

\section{QMUXING entangling scheme with channel loss}

The state of our system after the photon has interacted with Alice's
NV centers is given by Eq. (\ref{eq:3rd_ent_step}). Now we assume
that the photon travels across the channel and reaches Bob's side
with probability $P_{0}$ and it is lost with probability $1-P_{0}.$
Therefore, the final state will be described by the following density
matrix:

\noindent 
\begin{equation}
\begin{array}{c}
\rho_{tot}=P_{0}\rho_{1234}+(1-P_{0})(I_{13}\otimes\rho_{24}),\end{array}
\end{equation}
where $\rho_{1234}=|\psi_{1234}\left\rangle \left\langle \psi_{1234}|\right.\right.$
is the density matrix of the state of Eq. (\ref{eq:4th_ent_step}),
$I_{13}$ is the complete mixed state of the subspace spanned by the
base vectors $g_{1},$ $g_{3},$ $e_{1},$ and $e_{3}.$ $\rho_{24}=|\varphi_{24}\left\rangle \left\langle \varphi_{24}|\right.\right.$,
where $|\varphi_{24}\left\rangle =\left(|g_{2}\left\rangle +|e_{2}\left\rangle \right.\right.\right)\left(|g_{4}\left\rangle +|e_{4}\left\rangle \right.\right.\right)|0\left\rangle _{p}\right.\right.$
with $|0\left\rangle _{p}\right.$ the vacuum term of the photonic
mode. 

\section{\label{sec:deph_4qubits}Dephasing model in the four qubits QMUXING
entangling scheme applied to the Deutsch protocol}

Let us consider the situation where the quantum memories dephase over
time. The lost in fidelity during this time is $1-F$. For the sake
of simplicity we assume that Bob will measure a $D_{\mathcal{S}}$
photon (all the other cases can be obtained by flipping Bob's qubits).
A dephasing channel applied to the state of Eq. (\ref{eq:Ent_state_pur})
will produce the state.

\noindent 
\begin{equation}
\begin{array}{c}
\rho_{\mathrm{4}}=F^{2}\rho_{0}+F(1-F)(X_{1}\rho_{0}X_{1}+X_{3}\rho_{0}X_{3})\\
+(1-F)^{2}X_{1}X_{3}\rho_{0}X_{1}X_{3},
\end{array}
\end{equation}
where $\rho_{0}=|\psi_{0}\left\rangle \left\langle \psi_{0}|,\right.\right.$
and $|\psi_{0}\left\rangle =|\varphi_{12}^{+}\left\rangle |\varphi_{34}^{+}\left\rangle \right.\right.\right.$.
By following the Deutsch protocol Alice and Bob apply a CNOT gate
between QM3(QM4) and QM1(QM2), respectively. The resulting state will
be

\noindent 
\begin{equation}
\begin{array}{c}
\negthickspace\negthickspace\negthickspace\rho_{\mathrm{4}}=F^{2}\rho_{0}+F(1-F)(X_{1}\rho_{0}X_{1}+X_{1}X_{3}\rho_{0}X_{1}X_{3})\\
+(1-F)^{2}X_{3}\rho_{0}X_{3},
\end{array}
\end{equation}
Alice and Bob will measure their target qubits and they communicate
to each other the outcomes. The protocol is successful when the outcomes
are the same. If, for instance, the outcome are both $"+"$ the final
state will be given by

\noindent 
\begin{equation}
\begin{array}{c}
\rho_{\mathrm{4}}^{'}=\frac{F^{2}}{F^{2}+(1-F)^{2}}|\varphi_{34}^{+}\left\rangle \left\langle \varphi_{34}^{+}|\right.\right.\\
+\frac{(1-F)^{2}}{F^{2}+(1-F)^{2}}X_{3}|\varphi_{34}^{+}\left\rangle \left\langle \varphi_{34}^{+}|X_{3}.\right.\right.
\end{array}
\end{equation}

\bibliographystyle{apsrev4-1}
\bibliography{bib1}

\begin{thebibliography}{84}%
\makeatletter
\providecommand \@ifxundefined [1]{%
 \@ifx{#1\undefined}
}%
\providecommand \@ifnum [1]{%
 \ifnum #1\expandafter \@firstoftwo
 \else \expandafter \@secondoftwo
 \fi
}%
\providecommand \@ifx [1]{%
 \ifx #1\expandafter \@firstoftwo
 \else \expandafter \@secondoftwo
 \fi
}%
\providecommand \natexlab [1]{#1}%
\providecommand \enquote  [1]{``#1''}%
\providecommand \bibnamefont  [1]{#1}%
\providecommand \bibfnamefont [1]{#1}%
\providecommand \citenamefont [1]{#1}%
\providecommand \href@noop [0]{\@secondoftwo}%
\providecommand \href [0]{\begingroup \@sanitize@url \@href}%
\providecommand \@href[1]{\@@startlink{#1}\@@href}%
\providecommand \@@href[1]{\endgroup#1\@@endlink}%
\providecommand \@sanitize@url [0]{\catcode `\\12\catcode `\$12\catcode
  `\&12\catcode `\#12\catcode `\^12\catcode `\_12\catcode `\%12\relax}%
\providecommand \@@startlink[1]{}%
\providecommand \@@endlink[0]{}%
\providecommand \url  [0]{\begingroup\@sanitize@url \@url }%
\providecommand \@url [1]{\endgroup\@href {#1}{\urlprefix }}%
\providecommand \urlprefix  [0]{URL }%
\providecommand \Eprint [0]{\href }%
\providecommand \doibase [0]{http://dx.doi.org/}%
\providecommand \selectlanguage [0]{\@gobble}%
\providecommand \bibinfo  [0]{\@secondoftwo}%
\providecommand \bibfield  [0]{\@secondoftwo}%
\providecommand \translation [1]{[#1]}%
\providecommand \BibitemOpen [0]{}%
\providecommand \bibitemStop [0]{}%
\providecommand \bibitemNoStop [0]{.\EOS\space}%
\providecommand \EOS [0]{\spacefactor3000\relax}%
\providecommand \BibitemShut  [1]{\csname bibitem#1\endcsname}%
\let\auto@bib@innerbib\@empty
%</preamble>
\bibitem [{\citenamefont {Feynman}(1982)}]{feynman}%
  \BibitemOpen
  \bibfield  {author} {\bibinfo {author} {\bibfnamefont {R.~P.}\ \bibnamefont
  {Feynman}},\ }\href@noop {} {\bibfield  {journal} {\bibinfo  {journal}
  {International Journal of Theoretical Physics}\ }\textbf {\bibinfo {volume}
  {21}},\ \bibinfo {pages} {4467} (\bibinfo {year} {1982})}\BibitemShut
  {NoStop}%
\bibitem [{\citenamefont {Simon}(1994)}]{simon}%
  \BibitemOpen
  \bibfield  {author} {\bibinfo {author} {\bibfnamefont {D.~R.}\ \bibnamefont
  {Simon}},\ }\href@noop {} {\bibfield  {journal} {\bibinfo  {journal}
  {Foundations of Computer Science, 1994 Proceedings., 35th Annual Symposium
  on}\ ,\ \bibinfo {pages} {116}} (\bibinfo {year} {1994})}\BibitemShut
  {NoStop}%
\bibitem [{\citenamefont {DiVincenzo}(1995)}]{fDiVincenzo}%
  \BibitemOpen
  \bibfield  {author} {\bibinfo {author} {\bibfnamefont {D.~P.}\ \bibnamefont
  {DiVincenzo}},\ }\href@noop {} {\bibfield  {journal} {\bibinfo  {journal}
  {Science}\ }\textbf {\bibinfo {volume} {270}},\ \bibinfo {pages} {255}
  (\bibinfo {year} {1995})}\BibitemShut {NoStop}%
\bibitem [{\citenamefont {Dowling}\ and\ \citenamefont
  {Milburn}(2003)}]{Qtech}%
  \BibitemOpen
  \bibfield  {author} {\bibinfo {author} {\bibfnamefont {J.~P.}\ \bibnamefont
  {Dowling}}\ and\ \bibinfo {author} {\bibfnamefont {G.~J.}\ \bibnamefont
  {Milburn}},\ }\href@noop {} {\bibfield  {journal} {\bibinfo  {journal} {Phil.
  Trans. R. Soc. A}\ }\textbf {\bibinfo {volume} {361}},\ \bibinfo {pages}
  {3655} (\bibinfo {year} {2003})}\BibitemShut {NoStop}%
\bibitem [{\citenamefont {Dogen}\ \emph {et~al.}(2017)\citenamefont {Dogen},
  \citenamefont {Reinhard},\ and\ \citenamefont {Cappellaro}}]{QSensing}%
  \BibitemOpen
  \bibfield  {author} {\bibinfo {author} {\bibfnamefont {C.~L.}\ \bibnamefont
  {Dogen}}, \bibinfo {author} {\bibfnamefont {F.}~\bibnamefont {Reinhard}}, \
  and\ \bibinfo {author} {\bibfnamefont {P.}~\bibnamefont {Cappellaro}},\
  }\href@noop {} {\bibfield  {journal} {\bibinfo  {journal} {Rev. Mod. Phys.}\
  }\textbf {\bibinfo {volume} {89}},\ \bibinfo {pages} {035002} (\bibinfo
  {year} {2017})}\BibitemShut {NoStop}%
\bibitem [{\citenamefont {Simon}\ \emph {et~al.}(2014)\citenamefont {Simon},
  \citenamefont {Jaeger},\ and\ \citenamefont {Sergienko}}]{QImaging2}%
  \BibitemOpen
  \bibfield  {author} {\bibinfo {author} {\bibfnamefont {D.~S.}\ \bibnamefont
  {Simon}}, \bibinfo {author} {\bibfnamefont {G.}~\bibnamefont {Jaeger}}, \
  and\ \bibinfo {author} {\bibfnamefont {A.~V.}\ \bibnamefont {Sergienko}},\
  }\href@noop {} {\bibfield  {journal} {\bibinfo  {journal} {Int. J. Quantum
  Inform.}\ }\textbf {\bibinfo {volume} {12}},\ \bibinfo {pages} {1430004}
  (\bibinfo {year} {2014})}\BibitemShut {NoStop}%
\bibitem [{\citenamefont {Lugiato}\ \emph {et~al.}(2002)\citenamefont
  {Lugiato}, \citenamefont {gatti},\ and\ \citenamefont
  {Brambilla}}]{QImaging3}%
  \BibitemOpen
  \bibfield  {author} {\bibinfo {author} {\bibfnamefont {L.~A.}\ \bibnamefont
  {Lugiato}}, \bibinfo {author} {\bibfnamefont {A.}~\bibnamefont {gatti}}, \
  and\ \bibinfo {author} {\bibfnamefont {E.}~\bibnamefont {Brambilla}},\
  }\href@noop {} {\bibfield  {journal} {\bibinfo  {journal} {J. Opt. B}\
  }\textbf {\bibinfo {volume} {4}},\ \bibinfo {pages} {176} (\bibinfo {year}
  {2002})}\BibitemShut {NoStop}%
\bibitem [{\citenamefont {Bennett}\ and\ \citenamefont
  {Brassard}(2014)}]{QComm}%
  \BibitemOpen
  \bibfield  {author} {\bibinfo {author} {\bibfnamefont {C.~H.}\ \bibnamefont
  {Bennett}}\ and\ \bibinfo {author} {\bibfnamefont {G.}~\bibnamefont
  {Brassard}},\ }\href@noop {} {\bibfield  {journal} {\bibinfo  {journal}
  {Theoretical computer science}\ }\textbf {\bibinfo {volume} {560}},\ \bibinfo
  {pages} {7} (\bibinfo {year} {2014})}\BibitemShut {NoStop}%
\bibitem [{\citenamefont {Sangouard}\ \emph
  {et~al.}(2011{\natexlab{a}})\citenamefont {Sangouard}, \citenamefont {Simon},
  \citenamefont {De~Riedmatten},\ and\ \citenamefont {Gisin}}]{Quantumcomm}%
  \BibitemOpen
  \bibfield  {author} {\bibinfo {author} {\bibfnamefont {N.}~\bibnamefont
  {Sangouard}}, \bibinfo {author} {\bibfnamefont {C.}~\bibnamefont {Simon}},
  \bibinfo {author} {\bibfnamefont {C.}~\bibnamefont {De~Riedmatten}}, \ and\
  \bibinfo {author} {\bibfnamefont {N.}~\bibnamefont {Gisin}},\ }\href@noop {}
  {\bibfield  {journal} {\bibinfo  {journal} {Rev. Mod. Phys.}\ }\textbf
  {\bibinfo {volume} {83}},\ \bibinfo {pages} {33} (\bibinfo {year}
  {2011}{\natexlab{a}})}\BibitemShut {NoStop}%
\bibitem [{\citenamefont {Ekert}(1991{\natexlab{a}})}]{QKD01}%
  \BibitemOpen
  \bibfield  {author} {\bibinfo {author} {\bibfnamefont {A.~K.}\ \bibnamefont
  {Ekert}},\ }\href@noop {} {\bibfield  {journal} {\bibinfo  {journal} {Phys.
  Rev. Lett.}\ }\textbf {\bibinfo {volume} {67}},\ \bibinfo {pages} {661}
  (\bibinfo {year} {1991}{\natexlab{a}})}\BibitemShut {NoStop}%
\bibitem [{\citenamefont {Lo}(1999)}]{QKD02}%
  \BibitemOpen
  \bibfield  {author} {\bibinfo {author} {\bibfnamefont {H.}~\bibnamefont
  {Lo}},\ }\href@noop {} {\bibfield  {journal} {\bibinfo  {journal} {Science}\
  }\textbf {\bibinfo {volume} {283}},\ \bibinfo {pages} {2050} (\bibinfo {year}
  {1999})}\BibitemShut {NoStop}%
\bibitem [{\citenamefont {Hwang}(2003)}]{QKD03}%
  \BibitemOpen
  \bibfield  {author} {\bibinfo {author} {\bibfnamefont {W.-Y.}\ \bibnamefont
  {Hwang}},\ }\href@noop {} {\bibfield  {journal} {\bibinfo  {journal} {Phys.
  Rev. Lett.}\ }\textbf {\bibinfo {volume} {91}},\ \bibinfo {pages} {057901}
  (\bibinfo {year} {2003})}\BibitemShut {NoStop}%
\bibitem [{\citenamefont {Munro}\ \emph {et~al.}(2015)\citenamefont {Munro},
  \citenamefont {Azuma}, \citenamefont {Tamaki},\ and\ \citenamefont
  {Nemoto}}]{QC_Bill}%
  \BibitemOpen
  \bibfield  {author} {\bibinfo {author} {\bibfnamefont {W.~J.}\ \bibnamefont
  {Munro}}, \bibinfo {author} {\bibfnamefont {K.}~\bibnamefont {Azuma}},
  \bibinfo {author} {\bibfnamefont {K.}~\bibnamefont {Tamaki}}, \ and\ \bibinfo
  {author} {\bibfnamefont {K.}~\bibnamefont {Nemoto}},\ }\href@noop {}
  {\bibfield  {journal} {\bibinfo  {journal} {IEEE Journal of Selected Topics
  in Quantum Electronics}\ }\textbf {\bibinfo {volume} {21}},\ \bibinfo {pages}
  {6400813} (\bibinfo {year} {2015})}\BibitemShut {NoStop}%
\bibitem [{\citenamefont {Nielsen}\ and\ \citenamefont
  {Chuang}(2000)}]{Qcomp1}%
  \BibitemOpen
  \bibfield  {author} {\bibinfo {author} {\bibfnamefont {M.}~\bibnamefont
  {Nielsen}}\ and\ \bibinfo {author} {\bibfnamefont {I.}~\bibnamefont
  {Chuang}},\ }\href@noop {} {\emph {\bibinfo {title} {Quantum Computation and
  Quantum Information}}}\ (\bibinfo  {publisher} {Cambridge University Press,
  Cambridge},\ \bibinfo {year} {2000})\BibitemShut {NoStop}%
\bibitem [{\citenamefont {Bennett}\ and\ \citenamefont
  {DiVincenzo}(2000)}]{Qcomp2}%
  \BibitemOpen
  \bibfield  {author} {\bibinfo {author} {\bibfnamefont {C.}~\bibnamefont
  {Bennett}}\ and\ \bibinfo {author} {\bibfnamefont {D.}~\bibnamefont
  {DiVincenzo}},\ }\href@noop {} {\bibfield  {journal} {\bibinfo  {journal}
  {Nature}\ }\textbf {\bibinfo {volume} {404}},\ \bibinfo {pages} {247}
  (\bibinfo {year} {2000})}\BibitemShut {NoStop}%
\bibitem [{\citenamefont {Raussendorf}\ and\ \citenamefont
  {Briegel}(2001)}]{Quantum_comp2}%
  \BibitemOpen
  \bibfield  {author} {\bibinfo {author} {\bibfnamefont {R.}~\bibnamefont
  {Raussendorf}}\ and\ \bibinfo {author} {\bibfnamefont {H.~J.}\ \bibnamefont
  {Briegel}},\ }\href@noop {} {\bibfield  {journal} {\bibinfo  {journal} {Phys.
  Rev. Lett.}\ }\textbf {\bibinfo {volume} {86}},\ \bibinfo {pages} {5188}
  (\bibinfo {year} {2001})}\BibitemShut {NoStop}%
\bibitem [{\citenamefont {Knill}(2005{\natexlab{a}})}]{Quantum_comp3}%
  \BibitemOpen
  \bibfield  {author} {\bibinfo {author} {\bibfnamefont {E.}~\bibnamefont
  {Knill}},\ }\href@noop {} {\bibfield  {journal} {\bibinfo  {journal}
  {Nature}\ }\textbf {\bibinfo {volume} {434}},\ \bibinfo {pages} {39}
  (\bibinfo {year} {2005}{\natexlab{a}})}\BibitemShut {NoStop}%
\bibitem [{\citenamefont {Duan}\ and\ \citenamefont
  {Raussendorf}(2005)}]{Quantum_comp4}%
  \BibitemOpen
  \bibfield  {author} {\bibinfo {author} {\bibfnamefont {L.-M.}\ \bibnamefont
  {Duan}}\ and\ \bibinfo {author} {\bibfnamefont {R.}~\bibnamefont
  {Raussendorf}},\ }\href@noop {} {\bibfield  {journal} {\bibinfo  {journal}
  {Phys. Rev. Lett.}\ }\textbf {\bibinfo {volume} {95}},\ \bibinfo {pages}
  {080503} (\bibinfo {year} {2005})}\BibitemShut {NoStop}%
\bibitem [{\citenamefont {Devitt}\ \emph
  {et~al.}(2013{\natexlab{a}})\citenamefont {Devitt}, \citenamefont {Stephens},
  \citenamefont {Munro},\ and\ \citenamefont {Nemoto}}]{QComp_Bill}%
  \BibitemOpen
  \bibfield  {author} {\bibinfo {author} {\bibfnamefont {S.~J.}\ \bibnamefont
  {Devitt}}, \bibinfo {author} {\bibfnamefont {A.~M.}\ \bibnamefont
  {Stephens}}, \bibinfo {author} {\bibfnamefont {W.~J.}\ \bibnamefont {Munro}},
  \ and\ \bibinfo {author} {\bibfnamefont {K.}~\bibnamefont {Nemoto}},\
  }\href@noop {} {\bibfield  {journal} {\bibinfo  {journal} {Nature
  Communications}\ }\textbf {\bibinfo {volume} {4}},\ \bibinfo {pages} {2524}
  (\bibinfo {year} {2013}{\natexlab{a}})}\BibitemShut {NoStop}%
\bibitem [{\citenamefont {Ekert}(1991{\natexlab{b}})}]{QC1}%
  \BibitemOpen
  \bibfield  {author} {\bibinfo {author} {\bibfnamefont {A.}~\bibnamefont
  {Ekert}},\ }\href@noop {} {\bibfield  {journal} {\bibinfo  {journal} {Phys.
  Rev. Lett.}\ }\textbf {\bibinfo {volume} {67}},\ \bibinfo {pages} {661}
  (\bibinfo {year} {1991}{\natexlab{b}})}\BibitemShut {NoStop}%
\bibitem [{\citenamefont {Bennett}(1992)}]{QC2}%
  \BibitemOpen
  \bibfield  {author} {\bibinfo {author} {\bibfnamefont {C.}~\bibnamefont
  {Bennett}},\ }\href@noop {} {\bibfield  {journal} {\bibinfo  {journal}
  {Science}\ }\textbf {\bibinfo {volume} {257}},\ \bibinfo {pages} {752}
  (\bibinfo {year} {1992})}\BibitemShut {NoStop}%
\bibitem [{\citenamefont {Jennewein}\ \emph {et~al.}(2000)\citenamefont
  {Jennewein}, \citenamefont {Simon}, \citenamefont {Weihs}, \citenamefont
  {Weinfurter},\ and\ \citenamefont {Zeilinger}}]{QC3}%
  \BibitemOpen
  \bibfield  {author} {\bibinfo {author} {\bibfnamefont {T.}~\bibnamefont
  {Jennewein}}, \bibinfo {author} {\bibfnamefont {C.}~\bibnamefont {Simon}},
  \bibinfo {author} {\bibfnamefont {G.}~\bibnamefont {Weihs}}, \bibinfo
  {author} {\bibfnamefont {H.}~\bibnamefont {Weinfurter}}, \ and\ \bibinfo
  {author} {\bibfnamefont {A.}~\bibnamefont {Zeilinger}},\ }\href@noop {}
  {\bibfield  {journal} {\bibinfo  {journal} {Phys. Rev. Lett.}\ }\textbf
  {\bibinfo {volume} {84}},\ \bibinfo {pages} {4729} (\bibinfo {year}
  {2000})}\BibitemShut {NoStop}%
\bibitem [{\citenamefont {Long}\ and\ \citenamefont {Liu}(2002)}]{QC4}%
  \BibitemOpen
  \bibfield  {author} {\bibinfo {author} {\bibfnamefont {G.}~\bibnamefont
  {Long}}\ and\ \bibinfo {author} {\bibfnamefont {X.}~\bibnamefont {Liu}},\
  }\href@noop {} {\bibfield  {journal} {\bibinfo  {journal} {Phys. Rev. A}\
  }\textbf {\bibinfo {volume} {65}},\ \bibinfo {pages} {032302} (\bibinfo
  {year} {2002})}\BibitemShut {NoStop}%
\bibitem [{\citenamefont {Bennett}\ \emph
  {et~al.}(1996{\natexlab{a}})\citenamefont {Bennett}, \citenamefont
  {Brassard}, \citenamefont {Popesco}, \citenamefont {Schumacher},
  \citenamefont {Smolin},\ and\ \citenamefont {Wootters}}]{tel_1}%
  \BibitemOpen
  \bibfield  {author} {\bibinfo {author} {\bibfnamefont {C.}~\bibnamefont
  {Bennett}}, \bibinfo {author} {\bibfnamefont {G.}~\bibnamefont {Brassard}},
  \bibinfo {author} {\bibfnamefont {S.}~\bibnamefont {Popesco}}, \bibinfo
  {author} {\bibfnamefont {B.}~\bibnamefont {Schumacher}}, \bibinfo {author}
  {\bibfnamefont {J.}~\bibnamefont {Smolin}}, \ and\ \bibinfo {author}
  {\bibfnamefont {W.}~\bibnamefont {Wootters}},\ }\href@noop {} {\bibfield
  {journal} {\bibinfo  {journal} {Phys. Rev. Lett.}\ }\textbf {\bibinfo
  {volume} {76}},\ \bibinfo {pages} {722} (\bibinfo {year}
  {1996}{\natexlab{a}})}\BibitemShut {NoStop}%
\bibitem [{\citenamefont {Deng}\ \emph {et~al.}(2005)\citenamefont {Deng},
  \citenamefont {Li}, \citenamefont {Li}, \citenamefont {Zhou},\ and\
  \citenamefont {Wang}}]{tel_2}%
  \BibitemOpen
  \bibfield  {author} {\bibinfo {author} {\bibfnamefont {F.}~\bibnamefont
  {Deng}}, \bibinfo {author} {\bibfnamefont {C.}~\bibnamefont {Li}}, \bibinfo
  {author} {\bibfnamefont {Y.}~\bibnamefont {Li}}, \bibinfo {author}
  {\bibfnamefont {H.}~\bibnamefont {Zhou}}, \ and\ \bibinfo {author}
  {\bibfnamefont {Y.}~\bibnamefont {Wang}},\ }\href@noop {} {\bibfield
  {journal} {\bibinfo  {journal} {Phys. Rev. A}\ }\textbf {\bibinfo {volume}
  {72}},\ \bibinfo {pages} {022338} (\bibinfo {year} {2005})}\BibitemShut
  {NoStop}%
\bibitem [{\citenamefont {Barrett}\ \emph {et~al.}(2004)\citenamefont
  {Barrett}, \citenamefont {Chiaverini}, \citenamefont {Schaetz}, \citenamefont
  {Britton}, \citenamefont {Itano}, \citenamefont {Jost}, \citenamefont
  {Knill}, \citenamefont {Langer}, \citenamefont {Leibfried}, \citenamefont
  {Ozeri},\ and\ \citenamefont {Wineland}}]{Qtel2}%
  \BibitemOpen
  \bibfield  {author} {\bibinfo {author} {\bibfnamefont {M.~D.}\ \bibnamefont
  {Barrett}}, \bibinfo {author} {\bibfnamefont {J.}~\bibnamefont {Chiaverini}},
  \bibinfo {author} {\bibfnamefont {T.}~\bibnamefont {Schaetz}}, \bibinfo
  {author} {\bibfnamefont {J.}~\bibnamefont {Britton}}, \bibinfo {author}
  {\bibfnamefont {W.~M.}\ \bibnamefont {Itano}}, \bibinfo {author}
  {\bibfnamefont {J.~D.}\ \bibnamefont {Jost}}, \bibinfo {author}
  {\bibfnamefont {E.}~\bibnamefont {Knill}}, \bibinfo {author} {\bibfnamefont
  {C.}~\bibnamefont {Langer}}, \bibinfo {author} {\bibfnamefont
  {D.}~\bibnamefont {Leibfried}}, \bibinfo {author} {\bibfnamefont
  {R.}~\bibnamefont {Ozeri}}, \ and\ \bibinfo {author} {\bibfnamefont {D.~J.}\
  \bibnamefont {Wineland}},\ }\href@noop {} {\bibfield  {journal} {\bibinfo
  {journal} {Nature}\ }\textbf {\bibinfo {volume} {429}},\ \bibinfo {pages}
  {737} (\bibinfo {year} {2004})}\BibitemShut {NoStop}%
\bibitem [{\citenamefont {Ma}\ \emph {et~al.}(2012)\citenamefont {Ma},
  \citenamefont {Herbst}, \citenamefont {Scheidl}, \citenamefont {Wang},
  \citenamefont {Kropatschenk}, \citenamefont {Naylor}, \citenamefont
  {Wittmann}, \citenamefont {Mech}, \citenamefont {Kofler}, \citenamefont
  {Anisimova}, \citenamefont {Marakov}, \citenamefont {Jennewein},
  \citenamefont {Ursin},\ and\ \citenamefont {Zeilinger}}]{Qtel3}%
  \BibitemOpen
  \bibfield  {author} {\bibinfo {author} {\bibfnamefont {X.-S.}\ \bibnamefont
  {Ma}}, \bibinfo {author} {\bibfnamefont {T.}~\bibnamefont {Herbst}}, \bibinfo
  {author} {\bibfnamefont {T.}~\bibnamefont {Scheidl}}, \bibinfo {author}
  {\bibfnamefont {D.}~\bibnamefont {Wang}}, \bibinfo {author} {\bibfnamefont
  {S.}~\bibnamefont {Kropatschenk}}, \bibinfo {author} {\bibfnamefont
  {W.}~\bibnamefont {Naylor}}, \bibinfo {author} {\bibfnamefont
  {B.}~\bibnamefont {Wittmann}}, \bibinfo {author} {\bibfnamefont
  {A.}~\bibnamefont {Mech}}, \bibinfo {author} {\bibfnamefont {J.}~\bibnamefont
  {Kofler}}, \bibinfo {author} {\bibfnamefont {E.}~\bibnamefont {Anisimova}},
  \bibinfo {author} {\bibfnamefont {V.}~\bibnamefont {Marakov}}, \bibinfo
  {author} {\bibfnamefont {T.}~\bibnamefont {Jennewein}}, \bibinfo {author}
  {\bibfnamefont {T.}~\bibnamefont {Ursin}}, \ and\ \bibinfo {author}
  {\bibfnamefont {A.}~\bibnamefont {Zeilinger}},\ }\href@noop {} {\bibfield
  {journal} {\bibinfo  {journal} {Nature}\ }\textbf {\bibinfo {volume} {489}},\
  \bibinfo {pages} {269} (\bibinfo {year} {2012})}\BibitemShut {NoStop}%
\bibitem [{\citenamefont {Takesue}\ \emph {et~al.}(2015)\citenamefont
  {Takesue}, \citenamefont {Dyer}, \citenamefont {Martin}, \citenamefont
  {Verma}, \citenamefont {Mirin},\ and\ \citenamefont {Nam}}]{Qtel4}%
  \BibitemOpen
  \bibfield  {author} {\bibinfo {author} {\bibfnamefont {H.}~\bibnamefont
  {Takesue}}, \bibinfo {author} {\bibfnamefont {S.~D.}\ \bibnamefont {Dyer}},
  \bibinfo {author} {\bibfnamefont {J.~S.}\ \bibnamefont {Martin}}, \bibinfo
  {author} {\bibfnamefont {V.}~\bibnamefont {Verma}}, \bibinfo {author}
  {\bibfnamefont {R.~P.}\ \bibnamefont {Mirin}}, \ and\ \bibinfo {author}
  {\bibfnamefont {S.~W.}\ \bibnamefont {Nam}},\ }\href@noop {} {\bibfield
  {journal} {\bibinfo  {journal} {Optica}\ }\textbf {\bibinfo {volume} {2}},\
  \bibinfo {pages} {832} (\bibinfo {year} {2015})}\BibitemShut {NoStop}%
\bibitem [{\citenamefont {Horodecki}\ \emph {et~al.}(2007)\citenamefont
  {Horodecki}, \citenamefont {Horodecki}, \citenamefont {Horodecki},\ and\
  \citenamefont {Horodecki}}]{Qentang}%
  \BibitemOpen
  \bibfield  {author} {\bibinfo {author} {\bibfnamefont {R.}~\bibnamefont
  {Horodecki}}, \bibinfo {author} {\bibfnamefont {P.}~\bibnamefont
  {Horodecki}}, \bibinfo {author} {\bibfnamefont {M.}~\bibnamefont
  {Horodecki}}, \ and\ \bibinfo {author} {\bibfnamefont {K.}~\bibnamefont
  {Horodecki}},\ }\href@noop {} {\bibfield  {journal} {\bibinfo  {journal}
  {Rev. Mod. Phys.}\ }\textbf {\bibinfo {volume} {81}},\ \bibinfo {pages} {865}
  (\bibinfo {year} {2007})}\BibitemShut {NoStop}%
\bibitem [{\citenamefont {Mattle}\ \emph {et~al.}(1996)\citenamefont {Mattle},
  \citenamefont {Weinfurter}, \citenamefont {Kwiat},\ and\ \citenamefont
  {Zeilinger}}]{Qentang2}%
  \BibitemOpen
  \bibfield  {author} {\bibinfo {author} {\bibfnamefont {K.}~\bibnamefont
  {Mattle}}, \bibinfo {author} {\bibfnamefont {H.}~\bibnamefont {Weinfurter}},
  \bibinfo {author} {\bibfnamefont {P.~G.}\ \bibnamefont {Kwiat}}, \ and\
  \bibinfo {author} {\bibfnamefont {A.}~\bibnamefont {Zeilinger}},\ }\href@noop
  {} {\bibfield  {journal} {\bibinfo  {journal} {Phys. Rev. Lett.}\ }\textbf
  {\bibinfo {volume} {76}},\ \bibinfo {pages} {4656} (\bibinfo {year}
  {1996})}\BibitemShut {NoStop}%
\bibitem [{\citenamefont {Kwiat}\ \emph {et~al.}(1996)\citenamefont {Kwiat},
  \citenamefont {Mattle}, \citenamefont {Weinfurter},\ and\ \citenamefont
  {Zeilinger}}]{Qentang3}%
  \BibitemOpen
  \bibfield  {author} {\bibinfo {author} {\bibfnamefont {P.~G.}\ \bibnamefont
  {Kwiat}}, \bibinfo {author} {\bibfnamefont {K.}~\bibnamefont {Mattle}},
  \bibinfo {author} {\bibfnamefont {H.}~\bibnamefont {Weinfurter}}, \ and\
  \bibinfo {author} {\bibfnamefont {A.}~\bibnamefont {Zeilinger}},\ }\href@noop
  {} {\bibfield  {journal} {\bibinfo  {journal} {Optics and Photonics News}\
  }\textbf {\bibinfo {volume} {7}},\ \bibinfo {pages} {14} (\bibinfo {year}
  {1996})}\BibitemShut {NoStop}%
\bibitem [{\citenamefont {Vittorini}\ \emph {et~al.}()\citenamefont
  {Vittorini}, \citenamefont {Hucul}, \citenamefont {Inlek}, \citenamefont
  {Crocker},\ and\ \citenamefont {Monroe}}]{QMentanglement}%
  \BibitemOpen
  \bibfield  {author} {\bibinfo {author} {\bibfnamefont {G.}~\bibnamefont
  {Vittorini}}, \bibinfo {author} {\bibfnamefont {D.}~\bibnamefont {Hucul}},
  \bibinfo {author} {\bibfnamefont {V.}~\bibnamefont {Inlek}}, \bibinfo
  {author} {\bibfnamefont {C.}~\bibnamefont {Crocker}}, \ and\ \bibinfo
  {author} {\bibfnamefont {C.}~\bibnamefont {Monroe}},\ }\href@noop {}
  {\bibinfo  {journal} {Phys. Rev. A}\ }\BibitemShut {NoStop}%
\bibitem [{\citenamefont {Chou}\ \emph {et~al.}(2005)\citenamefont {Chou},
  \citenamefont {de~Riedmatten}, \citenamefont {Felinto}, \citenamefont
  {Polyakov}, \citenamefont {van Enk},\ and\ \citenamefont
  {Kimble}}]{QMentanglement2}%
  \BibitemOpen
\bibfield  {journal} {  }\bibfield  {author} {\bibinfo {author} {\bibfnamefont
  {C.}~\bibnamefont {Chou}}, \bibinfo {author} {\bibfnamefont {R.}~\bibnamefont
  {de~Riedmatten}}, \bibinfo {author} {\bibfnamefont {D.}~\bibnamefont
  {Felinto}}, \bibinfo {author} {\bibfnamefont {S.}~\bibnamefont {Polyakov}},
  \bibinfo {author} {\bibfnamefont {S.}~\bibnamefont {van Enk}}, \ and\
  \bibinfo {author} {\bibfnamefont {H.}~\bibnamefont {Kimble}},\ }\href@noop {}
  {\bibfield  {journal} {\bibinfo  {journal} {Nature (London)}\ }\textbf
  {\bibinfo {volume} {438}},\ \bibinfo {pages} {828} (\bibinfo {year}
  {2005})}\BibitemShut {NoStop}%
\bibitem [{\citenamefont {Moehring}\ \emph {et~al.}(2007)\citenamefont
  {Moehring}, \citenamefont {Maunz}, \citenamefont {Olmschenk}, \citenamefont
  {Younge}, \citenamefont {Matsukevich}, \citenamefont {Duan},\ and\
  \citenamefont {Monroe}}]{QMentanglement3}%
  \BibitemOpen
  \bibfield  {author} {\bibinfo {author} {\bibfnamefont {D.~L.}\ \bibnamefont
  {Moehring}}, \bibinfo {author} {\bibfnamefont {P.}~\bibnamefont {Maunz}},
  \bibinfo {author} {\bibfnamefont {S.}~\bibnamefont {Olmschenk}}, \bibinfo
  {author} {\bibfnamefont {K.}~\bibnamefont {Younge}}, \bibinfo {author}
  {\bibfnamefont {D.}~\bibnamefont {Matsukevich}}, \bibinfo {author}
  {\bibfnamefont {L.-M.}\ \bibnamefont {Duan}}, \ and\ \bibinfo {author}
  {\bibfnamefont {C.}~\bibnamefont {Monroe}},\ }\href@noop {} {\bibfield
  {journal} {\bibinfo  {journal} {Nature (London)}\ }\textbf {\bibinfo {volume}
  {449}},\ \bibinfo {pages} {68} (\bibinfo {year} {2007})}\BibitemShut
  {NoStop}%
\bibitem [{\citenamefont {Nolleke}\ \emph {et~al.}(2013)\citenamefont
  {Nolleke}, \citenamefont {Neuzner}, \citenamefont {Reiserer}, \citenamefont
  {Hahn}, \citenamefont {Rempe},\ and\ \citenamefont
  {Ritter}}]{QMentanglement4}%
  \BibitemOpen
  \bibfield  {author} {\bibinfo {author} {\bibfnamefont {C.}~\bibnamefont
  {Nolleke}}, \bibinfo {author} {\bibfnamefont {A.}~\bibnamefont {Neuzner}},
  \bibinfo {author} {\bibfnamefont {A.}~\bibnamefont {Reiserer}}, \bibinfo
  {author} {\bibfnamefont {C.}~\bibnamefont {Hahn}}, \bibinfo {author}
  {\bibfnamefont {G.}~\bibnamefont {Rempe}}, \ and\ \bibinfo {author}
  {\bibfnamefont {S.}~\bibnamefont {Ritter}},\ }\href@noop {} {\bibfield
  {journal} {\bibinfo  {journal} {Phys. Rev. Lett.}\ }\textbf {\bibinfo
  {volume} {110}},\ \bibinfo {pages} {140403} (\bibinfo {year}
  {2013})}\BibitemShut {NoStop}%
\bibitem [{\citenamefont {Bernien}\ \emph
  {et~al.}(2013{\natexlab{a}})\citenamefont {Bernien}, \citenamefont {Hensen},
  \citenamefont {Pfaff}, \citenamefont {Koolstra}, \citenamefont {Blok},
  \citenamefont {Robledo}, \citenamefont {Taminiau}, \citenamefont {Markham},
  \citenamefont {Twitchen}, \citenamefont {Childress},\ and\ \citenamefont
  {Hanson}}]{QMentanglement5}%
  \BibitemOpen
  \bibfield  {author} {\bibinfo {author} {\bibfnamefont {H.}~\bibnamefont
  {Bernien}}, \bibinfo {author} {\bibfnamefont {B.}~\bibnamefont {Hensen}},
  \bibinfo {author} {\bibfnamefont {W.}~\bibnamefont {Pfaff}}, \bibinfo
  {author} {\bibfnamefont {G.}~\bibnamefont {Koolstra}}, \bibinfo {author}
  {\bibfnamefont {M.}~\bibnamefont {Blok}}, \bibinfo {author} {\bibfnamefont
  {L.}~\bibnamefont {Robledo}}, \bibinfo {author} {\bibfnamefont
  {T.}~\bibnamefont {Taminiau}}, \bibinfo {author} {\bibfnamefont
  {M.}~\bibnamefont {Markham}}, \bibinfo {author} {\bibfnamefont
  {D.}~\bibnamefont {Twitchen}}, \bibinfo {author} {\bibfnamefont
  {L.}~\bibnamefont {Childress}}, \ and\ \bibinfo {author} {\bibfnamefont
  {R.}~\bibnamefont {Hanson}},\ }\href@noop {} {\bibfield  {journal} {\bibinfo
  {journal} {Nature (London)}\ }\textbf {\bibinfo {volume} {497}},\ \bibinfo
  {pages} {86} (\bibinfo {year} {2013}{\natexlab{a}})}\BibitemShut {NoStop}%
\bibitem [{\citenamefont {Razavi}\ \emph {et~al.}(2009)\citenamefont {Razavi},
  \citenamefont {Piani},\ and\ \citenamefont {Lutkenhaus}}]{multiple_mem_conf}%
  \BibitemOpen
  \bibfield  {author} {\bibinfo {author} {\bibfnamefont {M.}~\bibnamefont
  {Razavi}}, \bibinfo {author} {\bibfnamefont {M.}~\bibnamefont {Piani}}, \
  and\ \bibinfo {author} {\bibfnamefont {N.}~\bibnamefont {Lutkenhaus}},\
  }\href@noop {} {\bibfield  {journal} {\bibinfo  {journal} {Phys. Rev. A}\
  }\textbf {\bibinfo {volume} {80}},\ \bibinfo {pages} {032301} (\bibinfo
  {year} {2009})}\BibitemShut {NoStop}%
\bibitem [{\citenamefont {Collins}\ \emph {et~al.}(2007)\citenamefont
  {Collins}, \citenamefont {Jenkins}, \citenamefont {Kuzmich},\ and\
  \citenamefont {Kennedy}}]{MultiQM2}%
  \BibitemOpen
  \bibfield  {author} {\bibinfo {author} {\bibfnamefont {O.~A.}\ \bibnamefont
  {Collins}}, \bibinfo {author} {\bibfnamefont {S.~D.}\ \bibnamefont
  {Jenkins}}, \bibinfo {author} {\bibfnamefont {A.}~\bibnamefont {Kuzmich}}, \
  and\ \bibinfo {author} {\bibfnamefont {T.~A.~B.}\ \bibnamefont {Kennedy}},\
  }\href@noop {} {\bibfield  {journal} {\bibinfo  {journal} {Phys. Rev. Lett.}\
  }\textbf {\bibinfo {volume} {98}},\ \bibinfo {pages} {060502} (\bibinfo
  {year} {2007})}\BibitemShut {NoStop}%
\bibitem [{\citenamefont {Bennett}\ \emph
  {et~al.}(1996{\natexlab{b}})\citenamefont {Bennett}, \citenamefont
  {Brassard}, \citenamefont {Popescu}, \citenamefont {Schimacher},
  \citenamefont {Smolin},\ and\ \citenamefont {Wooters}}]{QP1}%
  \BibitemOpen
  \bibfield  {author} {\bibinfo {author} {\bibfnamefont {C.~H.}\ \bibnamefont
  {Bennett}}, \bibinfo {author} {\bibfnamefont {G.}~\bibnamefont {Brassard}},
  \bibinfo {author} {\bibfnamefont {S.}~\bibnamefont {Popescu}}, \bibinfo
  {author} {\bibfnamefont {B.}~\bibnamefont {Schimacher}}, \bibinfo {author}
  {\bibfnamefont {J.~A.}\ \bibnamefont {Smolin}}, \ and\ \bibinfo {author}
  {\bibfnamefont {W.~K.}\ \bibnamefont {Wooters}},\ }\href@noop {} {\bibfield
  {journal} {\bibinfo  {journal} {Phys. Rev. Lett.}\ }\textbf {\bibinfo
  {volume} {76}},\ \bibinfo {pages} {722} (\bibinfo {year}
  {1996}{\natexlab{b}})}\BibitemShut {NoStop}%
\bibitem [{\citenamefont {Deutsch}\ \emph {et~al.}(1996)\citenamefont
  {Deutsch}, \citenamefont {Ekert}, \citenamefont {Macchiavello}, \citenamefont
  {Popescu},\ and\ \citenamefont {Sanpera}}]{QP2Deutsch}%
  \BibitemOpen
  \bibfield  {author} {\bibinfo {author} {\bibfnamefont {D.}~\bibnamefont
  {Deutsch}}, \bibinfo {author} {\bibfnamefont {A.}~\bibnamefont {Ekert}},
  \bibinfo {author} {\bibfnamefont {C.}~\bibnamefont {Macchiavello}}, \bibinfo
  {author} {\bibfnamefont {S.}~\bibnamefont {Popescu}}, \ and\ \bibinfo
  {author} {\bibfnamefont {A.}~\bibnamefont {Sanpera}},\ }\href@noop {}
  {\bibfield  {journal} {\bibinfo  {journal} {Phys. Rev. Lett.}\ }\textbf
  {\bibinfo {volume} {77}},\ \bibinfo {pages} {2818} (\bibinfo {year}
  {1996})}\BibitemShut {NoStop}%
\bibitem [{\citenamefont {Dür}\ \emph {et~al.}(1999)\citenamefont {Dür},
  \citenamefont {Briegel}, \citenamefont {Cirac},\ and\ \citenamefont
  {Zoller}}]{QP3Dur}%
  \BibitemOpen
  \bibfield  {author} {\bibinfo {author} {\bibfnamefont {W.}~\bibnamefont
  {Dür}}, \bibinfo {author} {\bibfnamefont {H.-J.}\ \bibnamefont {Briegel}},
  \bibinfo {author} {\bibfnamefont {J.~I.}\ \bibnamefont {Cirac}}, \ and\
  \bibinfo {author} {\bibfnamefont {P.}~\bibnamefont {Zoller}},\ }\href@noop {}
  {\bibfield  {journal} {\bibinfo  {journal} {Phys. Rev. A}\ }\textbf {\bibinfo
  {volume} {59}},\ \bibinfo {pages} {169} (\bibinfo {year} {1999})}\BibitemShut
  {NoStop}%
\bibitem [{\citenamefont {Takeoka}\ \emph {et~al.}(2014)\citenamefont
  {Takeoka}, \citenamefont {Guha},\ and\ \citenamefont {Wilde}}]{Loss}%
  \BibitemOpen
  \bibfield  {author} {\bibinfo {author} {\bibfnamefont {M.}~\bibnamefont
  {Takeoka}}, \bibinfo {author} {\bibfnamefont {S.}~\bibnamefont {Guha}}, \
  and\ \bibinfo {author} {\bibfnamefont {M.~M.}\ \bibnamefont {Wilde}},\
  }\href@noop {} {\bibfield  {journal} {\bibinfo  {journal} {Nature Comm.}\
  }\textbf {\bibinfo {volume} {5}},\ \bibinfo {pages} {5235} (\bibinfo {year}
  {2014})}\BibitemShut {NoStop}%
\bibitem [{\citenamefont {Lo~Piparo}\ and\ \citenamefont
  {Razavi}(2015)}]{Loss2}%
  \BibitemOpen
  \bibfield  {author} {\bibinfo {author} {\bibfnamefont {N.}~\bibnamefont
  {Lo~Piparo}}\ and\ \bibinfo {author} {\bibfnamefont {M.}~\bibnamefont
  {Razavi}},\ }\href@noop {} {\bibfield  {journal} {\bibinfo  {journal} {IEEE
  Journal of Selected Topics in Quantum Electronics}\ }\textbf {\bibinfo
  {volume} {21}},\ \bibinfo {pages} {6601010} (\bibinfo {year}
  {2015})}\BibitemShut {NoStop}%
\bibitem [{\citenamefont {Togan}\ \emph
  {et~al.}(2010{\natexlab{a}})\citenamefont {Togan}, \citenamefont {Chu},
  \citenamefont {Trifonov}, \citenamefont {Jiang}, \citenamefont {Maze},
  \citenamefont {Childress}, \citenamefont {Dutt}, \citenamefont {Sorensen},
  \citenamefont {Hemmer},\ and\ \citenamefont {Zibrov}}]{QMent_creation}%
  \BibitemOpen
  \bibfield  {author} {\bibinfo {author} {\bibfnamefont {E.}~\bibnamefont
  {Togan}}, \bibinfo {author} {\bibfnamefont {Y.}~\bibnamefont {Chu}}, \bibinfo
  {author} {\bibfnamefont {A.~S.}\ \bibnamefont {Trifonov}}, \bibinfo {author}
  {\bibfnamefont {L.}~\bibnamefont {Jiang}}, \bibinfo {author} {\bibfnamefont
  {J.}~\bibnamefont {Maze}}, \bibinfo {author} {\bibfnamefont {L.}~\bibnamefont
  {Childress}}, \bibinfo {author} {\bibfnamefont {M.~V.}\ \bibnamefont {Dutt}},
  \bibinfo {author} {\bibfnamefont {A.~S.}\ \bibnamefont {Sorensen}}, \bibinfo
  {author} {\bibfnamefont {P.~R.}\ \bibnamefont {Hemmer}}, \ and\ \bibinfo
  {author} {\bibfnamefont {M.~D.}\ \bibnamefont {Zibrov}, \bibfnamefont
  {A.~S.and~Lukin}},\ }\href@noop {} {\bibfield  {journal} {\bibinfo  {journal}
  {Nature Letters}\ }\textbf {\bibinfo {volume} {466}},\ \bibinfo {pages} {730}
  (\bibinfo {year} {2010}{\natexlab{a}})}\BibitemShut {NoStop}%
\bibitem [{\citenamefont {Akopian}(2006)}]{QMent_emitters1}%
  \BibitemOpen
  \bibfield  {author} {\bibinfo {author} {\bibfnamefont {N.}~\bibnamefont
  {Akopian}},\ }\href@noop {} {\bibfield  {journal} {\bibinfo  {journal} {Phys.
  Rev. Lett.}\ }\textbf {\bibinfo {volume} {96}},\ \bibinfo {pages} {130501}
  (\bibinfo {year} {2006})}\BibitemShut {NoStop}%
\bibitem [{\citenamefont {Su}\ \emph {et~al.}(2008)\citenamefont {Su},
  \citenamefont {Greentree},\ and\ \citenamefont {Hollenberg}}]{NVemitter1}%
  \BibitemOpen
  \bibfield  {author} {\bibinfo {author} {\bibfnamefont {C.-H.}\ \bibnamefont
  {Su}}, \bibinfo {author} {\bibfnamefont {A.~D.}\ \bibnamefont {Greentree}}, \
  and\ \bibinfo {author} {\bibfnamefont {L.~C.~L.}\ \bibnamefont
  {Hollenberg}},\ }\href@noop {} {\bibfield  {journal} {\bibinfo  {journal}
  {Opt. Express}\ }\textbf {\bibinfo {volume} {16}},\ \bibinfo {pages} {6240}
  (\bibinfo {year} {2008})}\BibitemShut {NoStop}%
\bibitem [{\citenamefont {Saglamyurek}\ \emph {et~al.}(2011)\citenamefont
  {Saglamyurek}, \citenamefont {Sinclair}, \citenamefont {Jin}, \citenamefont
  {Slater}, \citenamefont {Oblak}, \citenamefont {Bussières}, \citenamefont
  {George}, \citenamefont {Ricken}, \citenamefont {W.},\ and\ \citenamefont
  {Tittel}}]{QMent_ab1}%
  \BibitemOpen
  \bibfield  {author} {\bibinfo {author} {\bibfnamefont {E.}~\bibnamefont
  {Saglamyurek}}, \bibinfo {author} {\bibfnamefont {N.}~\bibnamefont
  {Sinclair}}, \bibinfo {author} {\bibfnamefont {J.}~\bibnamefont {Jin}},
  \bibinfo {author} {\bibfnamefont {J.~A.}\ \bibnamefont {Slater}}, \bibinfo
  {author} {\bibfnamefont {D.}~\bibnamefont {Oblak}}, \bibinfo {author}
  {\bibfnamefont {F.}~\bibnamefont {Bussières}}, \bibinfo {author}
  {\bibfnamefont {M.}~\bibnamefont {George}}, \bibinfo {author} {\bibfnamefont
  {R.}~\bibnamefont {Ricken}}, \bibinfo {author} {\bibfnamefont
  {S.}~\bibnamefont {W.}}, \ and\ \bibinfo {author} {\bibfnamefont
  {W.}~\bibnamefont {Tittel}},\ }\href@noop {} {\bibfield  {journal} {\bibinfo
  {journal} {Nature}\ }\textbf {\bibinfo {volume} {469}},\ \bibinfo {pages}
  {512} (\bibinfo {year} {2011})}\BibitemShut {NoStop}%
\bibitem [{\citenamefont {Acosta}\ \emph {et~al.}(2010)\citenamefont {Acosta},
  \citenamefont {Bauch}, \citenamefont {Jarmola}, \citenamefont {Zipp},
  \citenamefont {Ledbetter},\ and\ \citenamefont {Budker}}]{NVabs1}%
  \BibitemOpen
  \bibfield  {author} {\bibinfo {author} {\bibfnamefont {V.~M.}\ \bibnamefont
  {Acosta}}, \bibinfo {author} {\bibfnamefont {E.}~\bibnamefont {Bauch}},
  \bibinfo {author} {\bibfnamefont {A.}~\bibnamefont {Jarmola}}, \bibinfo
  {author} {\bibfnamefont {L.~J.}\ \bibnamefont {Zipp}}, \bibinfo {author}
  {\bibfnamefont {M.~P.}\ \bibnamefont {Ledbetter}}, \ and\ \bibinfo {author}
  {\bibfnamefont {D.}~\bibnamefont {Budker}},\ }\href@noop {} {\bibfield
  {journal} {\bibinfo  {journal} {Appl. Phys. Lett.}\ }\textbf {\bibinfo
  {volume} {97}},\ \bibinfo {pages} {174104} (\bibinfo {year}
  {2010})}\BibitemShut {NoStop}%
\bibitem [{\citenamefont {Nemoto}\ \emph {et~al.}(2014)\citenamefont {Nemoto},
  \citenamefont {Trupke}, \citenamefont {Devitt}, \citenamefont {Stephens},
  \citenamefont {Sharfenberger}, \citenamefont {Buczak}, \citenamefont
  {Nobauer}, \citenamefont {T.}, \citenamefont {Everitt}, \citenamefont
  {Schmiedmayer},\ and\ \citenamefont {Munro}}]{QM_refl1}%
  \BibitemOpen
  \bibfield  {author} {\bibinfo {author} {\bibfnamefont {K.}~\bibnamefont
  {Nemoto}}, \bibinfo {author} {\bibfnamefont {M.}~\bibnamefont {Trupke}},
  \bibinfo {author} {\bibfnamefont {S.~J.}\ \bibnamefont {Devitt}}, \bibinfo
  {author} {\bibfnamefont {A.~M.}\ \bibnamefont {Stephens}}, \bibinfo {author}
  {\bibfnamefont {B.}~\bibnamefont {Sharfenberger}}, \bibinfo {author}
  {\bibfnamefont {K.}~\bibnamefont {Buczak}}, \bibinfo {author} {\bibfnamefont
  {T.}~\bibnamefont {Nobauer}}, \bibinfo {author} {\bibnamefont {T.}}, \bibinfo
  {author} {\bibfnamefont {M.~S.}\ \bibnamefont {Everitt}}, \bibinfo {author}
  {\bibfnamefont {J.}~\bibnamefont {Schmiedmayer}}, \ and\ \bibinfo {author}
  {\bibfnamefont {W.~J.}\ \bibnamefont {Munro}},\ }\href@noop {} {\bibfield
  {journal} {\bibinfo  {journal} {Phys. Rev. X}\ }\textbf {\bibinfo {volume}
  {4}},\ \bibinfo {pages} {031022} (\bibinfo {year} {2014})}\BibitemShut
  {NoStop}%
\bibitem [{\citenamefont {Hu}\ \emph {et~al.}(2008)\citenamefont {Hu},
  \citenamefont {Young}, \citenamefont {O'Brien}, \citenamefont {Munro},\ and\
  \citenamefont {Rarity}}]{NVreflect}%
  \BibitemOpen
  \bibfield  {author} {\bibinfo {author} {\bibfnamefont {C.~H.}\ \bibnamefont
  {Hu}}, \bibinfo {author} {\bibfnamefont {A.}~\bibnamefont {Young}}, \bibinfo
  {author} {\bibfnamefont {J.~L.}\ \bibnamefont {O'Brien}}, \bibinfo {author}
  {\bibfnamefont {W.~J.}\ \bibnamefont {Munro}}, \ and\ \bibinfo {author}
  {\bibfnamefont {J.~G.}\ \bibnamefont {Rarity}},\ }\href@noop {} {\bibfield
  {journal} {\bibinfo  {journal} {Phys. Rev. B}\ }\textbf {\bibinfo {volume}
  {78}},\ \bibinfo {pages} {085307} (\bibinfo {year} {2008})}\BibitemShut
  {NoStop}%
\bibitem [{\citenamefont {Harty}\ \emph {et~al.}(2014)\citenamefont {Harty},
  \citenamefont {Allcock}, \citenamefont {Ballance}, \citenamefont {Guidoni},
  \citenamefont {Janacek}, \citenamefont {Linke}, \citenamefont {Stacey},\ and\
  \citenamefont {Lucas}}]{trapped_ions}%
  \BibitemOpen
  \bibfield  {author} {\bibinfo {author} {\bibfnamefont {T.~P.}\ \bibnamefont
  {Harty}}, \bibinfo {author} {\bibfnamefont {D.~T.~C.}\ \bibnamefont
  {Allcock}}, \bibinfo {author} {\bibfnamefont {C.~J.}\ \bibnamefont
  {Ballance}}, \bibinfo {author} {\bibfnamefont {L.}~\bibnamefont {Guidoni}},
  \bibinfo {author} {\bibfnamefont {H.~A.}\ \bibnamefont {Janacek}}, \bibinfo
  {author} {\bibfnamefont {N.~M.}\ \bibnamefont {Linke}}, \bibinfo {author}
  {\bibfnamefont {D.~N.}\ \bibnamefont {Stacey}}, \ and\ \bibinfo {author}
  {\bibfnamefont {D.~M.}\ \bibnamefont {Lucas}},\ }\href@noop {} {\bibfield
  {journal} {\bibinfo  {journal} {Phys. Rev. Lett.}\ }\textbf {\bibinfo
  {volume} {113}},\ \bibinfo {pages} {220501} (\bibinfo {year}
  {2014})}\BibitemShut {NoStop}%
\bibitem [{\citenamefont {Bohnet}\ \emph {et~al.}(2016)\citenamefont {Bohnet},
  \citenamefont {Sawyer}, \citenamefont {Britton}, \citenamefont {Wall},
  \citenamefont {Rey}, \citenamefont {Foss-Feig},\ and\ \citenamefont
  {Bollinger}}]{ionTrap1}%
  \BibitemOpen
  \bibfield  {author} {\bibinfo {author} {\bibfnamefont {J.~G.}\ \bibnamefont
  {Bohnet}}, \bibinfo {author} {\bibfnamefont {B.~C.}\ \bibnamefont {Sawyer}},
  \bibinfo {author} {\bibfnamefont {J.~W.}\ \bibnamefont {Britton}}, \bibinfo
  {author} {\bibfnamefont {M.~L.}\ \bibnamefont {Wall}}, \bibinfo {author}
  {\bibfnamefont {A.~M.}\ \bibnamefont {Rey}}, \bibinfo {author} {\bibfnamefont
  {M.}~\bibnamefont {Foss-Feig}}, \ and\ \bibinfo {author} {\bibfnamefont
  {J.~J.}\ \bibnamefont {Bollinger}},\ }\href@noop {} {\bibfield  {journal}
  {\bibinfo  {journal} {Science}\ }\textbf {\bibinfo {volume} {352}},\ \bibinfo
  {pages} {1297} (\bibinfo {year} {2016})}\BibitemShut {NoStop}%
\bibitem [{\citenamefont {Kalb}\ \emph {et~al.}(2015)\citenamefont {Kalb},
  \citenamefont {Reiserer}, \citenamefont {Ritter},\ and\ \citenamefont
  {Rempe}}]{trapped_at}%
  \BibitemOpen
  \bibfield  {author} {\bibinfo {author} {\bibfnamefont {N.}~\bibnamefont
  {Kalb}}, \bibinfo {author} {\bibfnamefont {A.}~\bibnamefont {Reiserer}},
  \bibinfo {author} {\bibfnamefont {S.}~\bibnamefont {Ritter}}, \ and\ \bibinfo
  {author} {\bibfnamefont {G.}~\bibnamefont {Rempe}},\ }\href@noop {}
  {\bibfield  {journal} {\bibinfo  {journal} {Phys. Rev. Lett.}\ }\textbf
  {\bibinfo {volume} {114}},\ \bibinfo {pages} {220501} (\bibinfo {year}
  {2015})}\BibitemShut {NoStop}%
\bibitem [{\citenamefont {Blinov}\ \emph {et~al.}(2004)\citenamefont {Blinov},
  \citenamefont {Moehring}, \citenamefont {Duan},\ and\ \citenamefont
  {Monrow}}]{trapAtoms2}%
  \BibitemOpen
  \bibfield  {author} {\bibinfo {author} {\bibfnamefont {B.~B.}\ \bibnamefont
  {Blinov}}, \bibinfo {author} {\bibfnamefont {D.~L.}\ \bibnamefont
  {Moehring}}, \bibinfo {author} {\bibfnamefont {L.-M.}\ \bibnamefont {Duan}},
  \ and\ \bibinfo {author} {\bibfnamefont {C.}~\bibnamefont {Monrow}},\
  }\href@noop {} {\bibfield  {journal} {\bibinfo  {journal} {Nature}\ }\textbf
  {\bibinfo {volume} {428}},\ \bibinfo {pages} {153} (\bibinfo {year}
  {2004})}\BibitemShut {NoStop}%
\bibitem [{\citenamefont {Greve}\ \emph {et~al.}(2012)\citenamefont {Greve},
  \citenamefont {Yu}, \citenamefont {McMahon}, \citenamefont {Pelc},
  \citenamefont {Natarajan}, \citenamefont {Kim}, \citenamefont {Abe},
  \citenamefont {Maier}, \citenamefont {Schneider}, \citenamefont {Kamp},
  \citenamefont {Hofling}, \citenamefont {Hadfield}, \citenamefont {Forchel},
  \citenamefont {Fejer},\ and\ \citenamefont {Yamamoto}}]{qdots}%
  \BibitemOpen
  \bibfield  {author} {\bibinfo {author} {\bibfnamefont {K.~D.}\ \bibnamefont
  {Greve}}, \bibinfo {author} {\bibfnamefont {L.}~\bibnamefont {Yu}}, \bibinfo
  {author} {\bibfnamefont {P.~L.}\ \bibnamefont {McMahon}}, \bibinfo {author}
  {\bibfnamefont {J.~S.}\ \bibnamefont {Pelc}}, \bibinfo {author}
  {\bibfnamefont {C.~M.}\ \bibnamefont {Natarajan}}, \bibinfo {author}
  {\bibfnamefont {N.~Y.}\ \bibnamefont {Kim}}, \bibinfo {author} {\bibfnamefont
  {E.}~\bibnamefont {Abe}}, \bibinfo {author} {\bibfnamefont {S.}~\bibnamefont
  {Maier}}, \bibinfo {author} {\bibfnamefont {C.}~\bibnamefont {Schneider}},
  \bibinfo {author} {\bibfnamefont {M.}~\bibnamefont {Kamp}}, \bibinfo {author}
  {\bibfnamefont {S.}~\bibnamefont {Hofling}}, \bibinfo {author} {\bibfnamefont
  {R.~H.}\ \bibnamefont {Hadfield}}, \bibinfo {author} {\bibfnamefont
  {A.}~\bibnamefont {Forchel}}, \bibinfo {author} {\bibfnamefont {M.~M.}\
  \bibnamefont {Fejer}}, \ and\ \bibinfo {author} {\bibfnamefont
  {Y.}~\bibnamefont {Yamamoto}},\ }\href@noop {} {\bibfield  {journal}
  {\bibinfo  {journal} {Nature}\ }\textbf {\bibinfo {volume} {491}},\ \bibinfo
  {pages} {421} (\bibinfo {year} {2012})}\BibitemShut {NoStop}%
\bibitem [{\citenamefont {Huber}\ \emph {et~al.}(2017)\citenamefont {Huber},
  \citenamefont {Reindl}, \citenamefont {Huo}, \citenamefont {Huang},
  \citenamefont {Wildmann}, \citenamefont {Schmidt}, \citenamefont {Rastelli},\
  and\ \citenamefont {Trotta}}]{Qdots_ent2}%
  \BibitemOpen
  \bibfield  {author} {\bibinfo {author} {\bibfnamefont {D.}~\bibnamefont
  {Huber}}, \bibinfo {author} {\bibfnamefont {M.}~\bibnamefont {Reindl}},
  \bibinfo {author} {\bibfnamefont {Y.}~\bibnamefont {Huo}}, \bibinfo {author}
  {\bibfnamefont {H.}~\bibnamefont {Huang}}, \bibinfo {author} {\bibfnamefont
  {J.~S.}\ \bibnamefont {Wildmann}}, \bibinfo {author} {\bibfnamefont {O.~G.}\
  \bibnamefont {Schmidt}}, \bibinfo {author} {\bibfnamefont {A.}~\bibnamefont
  {Rastelli}}, \ and\ \bibinfo {author} {\bibfnamefont {R.}~\bibnamefont
  {Trotta}},\ }\href@noop {} {\bibfield  {journal} {\bibinfo  {journal} {Nature
  Comm.}\ }\textbf {\bibinfo {volume} {8}} (\bibinfo {year}
  {2017})}\BibitemShut {NoStop}%
\bibitem [{\citenamefont {Gao}\ \emph {et~al.}(2012)\citenamefont {Gao},
  \citenamefont {Fallahi}, \citenamefont {Togan}, \citenamefont
  {Miguel-Sanchez},\ and\ \citenamefont {Imamoglu}}]{Qdots_ent3}%
  \BibitemOpen
  \bibfield  {author} {\bibinfo {author} {\bibfnamefont {W.~B.}\ \bibnamefont
  {Gao}}, \bibinfo {author} {\bibfnamefont {P.}~\bibnamefont {Fallahi}},
  \bibinfo {author} {\bibfnamefont {E.}~\bibnamefont {Togan}}, \bibinfo
  {author} {\bibfnamefont {J.}~\bibnamefont {Miguel-Sanchez}}, \ and\ \bibinfo
  {author} {\bibfnamefont {A.}~\bibnamefont {Imamoglu}},\ }\href@noop {}
  {\bibfield  {journal} {\bibinfo  {journal} {Nature}\ }\textbf {\bibinfo
  {volume} {491}},\ \bibinfo {pages} {426} (\bibinfo {year}
  {2012})}\BibitemShut {NoStop}%
\bibitem [{\citenamefont {Liu}\ \emph {et~al.}(2013)\citenamefont {Liu},
  \citenamefont {Yu}, \citenamefont {Li},\ and\ \citenamefont
  {Wu}}]{NVcenters1}%
  \BibitemOpen
  \bibfield  {author} {\bibinfo {author} {\bibfnamefont {S.}~\bibnamefont
  {Liu}}, \bibinfo {author} {\bibfnamefont {R.}~\bibnamefont {Yu}}, \bibinfo
  {author} {\bibfnamefont {J.}~\bibnamefont {Li}}, \ and\ \bibinfo {author}
  {\bibfnamefont {Y.}~\bibnamefont {Wu}},\ }\href@noop {} {\bibfield  {journal}
  {\bibinfo  {journal} {Journal of Applied Physics}\ }\textbf {\bibinfo
  {volume} {114}},\ \bibinfo {pages} {244306} (\bibinfo {year}
  {2013})}\BibitemShut {NoStop}%
\bibitem [{\citenamefont {Liu}\ \emph {et~al.}(2015)\citenamefont {Liu},
  \citenamefont {Chen}, \citenamefont {ROng}, \citenamefont {P.}, \citenamefont
  {Jelezko}, \citenamefont {Tamura}, \citenamefont {Tanii}, \citenamefont
  {Teraji}, \citenamefont {Onoda}, \citenamefont {Isoya}, \citenamefont
  {Shinada}, \citenamefont {Wu},\ and\ \citenamefont {Zeng}}]{SiV_center}%
  \BibitemOpen
  \bibfield  {author} {\bibinfo {author} {\bibfnamefont {Y.}~\bibnamefont
  {Liu}}, \bibinfo {author} {\bibfnamefont {G.}~\bibnamefont {Chen}}, \bibinfo
  {author} {\bibfnamefont {Y.}~\bibnamefont {ROng}}, \bibinfo {author}
  {\bibfnamefont {M.~L.}\ \bibnamefont {P.}}, \bibinfo {author} {\bibfnamefont
  {F.}~\bibnamefont {Jelezko}}, \bibinfo {author} {\bibfnamefont
  {S.}~\bibnamefont {Tamura}}, \bibinfo {author} {\bibfnamefont
  {T.}~\bibnamefont {Tanii}}, \bibinfo {author} {\bibfnamefont
  {T.}~\bibnamefont {Teraji}}, \bibinfo {author} {\bibfnamefont
  {T.}~\bibnamefont {Onoda}, \bibfnamefont {S.~Ohshima}}, \bibinfo {author}
  {\bibfnamefont {J.}~\bibnamefont {Isoya}}, \bibinfo {author} {\bibfnamefont
  {T.}~\bibnamefont {Shinada}}, \bibinfo {author} {\bibfnamefont
  {E.}~\bibnamefont {Wu}}, \ and\ \bibinfo {author} {\bibfnamefont
  {H.}~\bibnamefont {Zeng}},\ }\href@noop {} {\bibfield  {journal} {\bibinfo
  {journal} {Scientific Reports}\ ,\ \bibinfo {pages} {12244}} (\bibinfo {year}
  {2015})}\BibitemShut {NoStop}%
\bibitem [{\citenamefont {Togan}\ \emph
  {et~al.}(2010{\natexlab{b}})\citenamefont {Togan}, \citenamefont {Chu},
  \citenamefont {Trifonov}, \citenamefont {Jiang}, \citenamefont {Maze},
  \citenamefont {Childress}, \citenamefont {Dutt}, \citenamefont {Sorensen},
  \citenamefont {Hemmer}, \citenamefont {Zibrov},\ and\ \citenamefont
  {Lukin}}]{NVexp1}%
  \BibitemOpen
  \bibfield  {author} {\bibinfo {author} {\bibfnamefont {E.}~\bibnamefont
  {Togan}}, \bibinfo {author} {\bibfnamefont {Y.}~\bibnamefont {Chu}}, \bibinfo
  {author} {\bibfnamefont {A.~S.}\ \bibnamefont {Trifonov}}, \bibinfo {author}
  {\bibfnamefont {L.}~\bibnamefont {Jiang}}, \bibinfo {author} {\bibfnamefont
  {J.}~\bibnamefont {Maze}}, \bibinfo {author} {\bibfnamefont {L.}~\bibnamefont
  {Childress}}, \bibinfo {author} {\bibfnamefont {M.~V.}\ \bibnamefont {Dutt}},
  \bibinfo {author} {\bibfnamefont {A.~S.}\ \bibnamefont {Sorensen}}, \bibinfo
  {author} {\bibfnamefont {P.~R.}\ \bibnamefont {Hemmer}}, \bibinfo {author}
  {\bibfnamefont {A.~S.}\ \bibnamefont {Zibrov}}, \ and\ \bibinfo {author}
  {\bibfnamefont {M.~D.}\ \bibnamefont {Lukin}},\ }\href@noop {} {\bibfield
  {journal} {\bibinfo  {journal} {Nature Letters}\ }\textbf {\bibinfo {volume}
  {466}},\ \bibinfo {pages} {730} (\bibinfo {year}
  {2010}{\natexlab{b}})}\BibitemShut {NoStop}%
\bibitem [{\citenamefont {Bernien}\ \emph
  {et~al.}(2013{\natexlab{b}})\citenamefont {Bernien}, \citenamefont {Hensen},
  \citenamefont {Pfaff}, \citenamefont {Koolstra}, \citenamefont {Blok},
  \citenamefont {Robledo}, \citenamefont {Taminiau}, \citenamefont {Markham},
  \citenamefont {Twitchen}, \citenamefont {Childress},\ and\ \citenamefont
  {Hanson}}]{NVexp2}%
  \BibitemOpen
  \bibfield  {author} {\bibinfo {author} {\bibfnamefont {H.}~\bibnamefont
  {Bernien}}, \bibinfo {author} {\bibfnamefont {B.}~\bibnamefont {Hensen}},
  \bibinfo {author} {\bibfnamefont {W.}~\bibnamefont {Pfaff}}, \bibinfo
  {author} {\bibfnamefont {G.}~\bibnamefont {Koolstra}}, \bibinfo {author}
  {\bibfnamefont {M.~S.}\ \bibnamefont {Blok}}, \bibinfo {author}
  {\bibfnamefont {L.}~\bibnamefont {Robledo}}, \bibinfo {author} {\bibfnamefont
  {T.~H.}\ \bibnamefont {Taminiau}}, \bibinfo {author} {\bibfnamefont
  {M.}~\bibnamefont {Markham}}, \bibinfo {author} {\bibfnamefont {D.~J.}\
  \bibnamefont {Twitchen}}, \bibinfo {author} {\bibfnamefont {L.}~\bibnamefont
  {Childress}}, \ and\ \bibinfo {author} {\bibfnamefont {R.}~\bibnamefont
  {Hanson}},\ }\href@noop {} {\bibfield  {journal} {\bibinfo  {journal}
  {Nature}\ }\textbf {\bibinfo {volume} {497}},\ \bibinfo {pages} {86}
  (\bibinfo {year} {2013}{\natexlab{b}})}\BibitemShut {NoStop}%
\bibitem [{\citenamefont {Albrecht}\ \emph {et~al.}(2013)\citenamefont
  {Albrecht}, \citenamefont {Bommer}, \citenamefont {Deutsch}, \citenamefont
  {Reichel},\ and\ \citenamefont {Becher}}]{NVcavity1}%
  \BibitemOpen
  \bibfield  {author} {\bibinfo {author} {\bibfnamefont {R.}~\bibnamefont
  {Albrecht}}, \bibinfo {author} {\bibfnamefont {A.}~\bibnamefont {Bommer}},
  \bibinfo {author} {\bibfnamefont {C.}~\bibnamefont {Deutsch}}, \bibinfo
  {author} {\bibfnamefont {J.}~\bibnamefont {Reichel}}, \ and\ \bibinfo
  {author} {\bibfnamefont {C.}~\bibnamefont {Becher}},\ }\href@noop {}
  {\bibfield  {journal} {\bibinfo  {journal} {Phys. Rev. Lett.}\ }\textbf
  {\bibinfo {volume} {110}},\ \bibinfo {pages} {243602} (\bibinfo {year}
  {2013})}\BibitemShut {NoStop}%
\bibitem [{\citenamefont {Lo~Piparo}\ \emph
  {et~al.}(2017{\natexlab{a}})\citenamefont {Lo~Piparo}, \citenamefont
  {Razavi},\ and\ \citenamefont {Munro}}]{DE}%
  \BibitemOpen
  \bibfield  {author} {\bibinfo {author} {\bibfnamefont {N.}~\bibnamefont
  {Lo~Piparo}}, \bibinfo {author} {\bibfnamefont {M.}~\bibnamefont {Razavi}}, \
  and\ \bibinfo {author} {\bibfnamefont {W.~J.}\ \bibnamefont {Munro}},\
  }\href@noop {} {\bibfield  {journal} {\bibinfo  {journal} {Phys. Rev. A}\
  }\textbf {\bibinfo {volume} {95}},\ \bibinfo {pages} {022338} (\bibinfo
  {year} {2017}{\natexlab{a}})}\BibitemShut {NoStop}%
\bibitem [{\citenamefont {Lo~Piparo}\ \emph
  {et~al.}(2017{\natexlab{b}})\citenamefont {Lo~Piparo}, \citenamefont
  {Razavi},\ and\ \citenamefont {Munro}}]{singleNVQKD}%
  \BibitemOpen
  \bibfield  {author} {\bibinfo {author} {\bibfnamefont {N.}~\bibnamefont
  {Lo~Piparo}}, \bibinfo {author} {\bibfnamefont {M.}~\bibnamefont {Razavi}}, \
  and\ \bibinfo {author} {\bibfnamefont {W.~J.}\ \bibnamefont {Munro}},\
  }\href@noop {} {\bibfield  {journal} {\bibinfo  {journal} {Phys. Rev. A}\
  }\textbf {\bibinfo {volume} {96}},\ \bibinfo {pages} {052313} (\bibinfo
  {year} {2017}{\natexlab{b}})}\BibitemShut {NoStop}%
\bibitem [{\citenamefont {Nemoto}\ \emph {et~al.}(2016)\citenamefont {Nemoto},
  \citenamefont {Trupke}, \citenamefont {Devitt}, \citenamefont
  {Sharfenberger}, \citenamefont {Buczak}, \citenamefont {Schmiedmayer},\ and\
  \citenamefont {Munro}}]{NVKae}%
  \BibitemOpen
  \bibfield  {author} {\bibinfo {author} {\bibfnamefont {K.}~\bibnamefont
  {Nemoto}}, \bibinfo {author} {\bibfnamefont {M.}~\bibnamefont {Trupke}},
  \bibinfo {author} {\bibfnamefont {S.~J.}\ \bibnamefont {Devitt}}, \bibinfo
  {author} {\bibfnamefont {B.}~\bibnamefont {Sharfenberger}}, \bibinfo {author}
  {\bibfnamefont {K.}~\bibnamefont {Buczak}}, \bibinfo {author} {\bibfnamefont
  {J.}~\bibnamefont {Schmiedmayer}}, \ and\ \bibinfo {author} {\bibfnamefont
  {W.~J.}\ \bibnamefont {Munro}},\ }\href@noop {} {\bibfield  {journal}
  {\bibinfo  {journal} {Scientific Reports}\ }\textbf {\bibinfo {volume} {6}},\
  \bibinfo {pages} {26284} (\bibinfo {year} {2016})}\BibitemShut {NoStop}%
\bibitem [{\citenamefont {Childress}\ and\ \citenamefont
  {Hanson}(2013)}]{NVnetworks1}%
  \BibitemOpen
  \bibfield  {author} {\bibinfo {author} {\bibfnamefont {L.}~\bibnamefont
  {Childress}}\ and\ \bibinfo {author} {\bibfnamefont {R.}~\bibnamefont
  {Hanson}},\ }\href@noop {} {\bibfield  {journal} {\bibinfo  {journal} {MRS
  Bulletin}\ }\textbf {\bibinfo {volume} {38}},\ \bibinfo {pages} {134}
  (\bibinfo {year} {2013})}\BibitemShut {NoStop}%
\bibitem [{\citenamefont {Blok}\ \emph {et~al.}(2015)\citenamefont {Blok},
  \citenamefont {Kalb}, \citenamefont {Reiserer}, \citenamefont {Taminiau},\
  and\ \citenamefont {Hanson}}]{NVnetworks2}%
  \BibitemOpen
  \bibfield  {author} {\bibinfo {author} {\bibfnamefont {M.~S.}\ \bibnamefont
  {Blok}}, \bibinfo {author} {\bibfnamefont {N.}~\bibnamefont {Kalb}}, \bibinfo
  {author} {\bibfnamefont {A.}~\bibnamefont {Reiserer}}, \bibinfo {author}
  {\bibfnamefont {T.~H.}\ \bibnamefont {Taminiau}}, \ and\ \bibinfo {author}
  {\bibfnamefont {R.}~\bibnamefont {Hanson}},\ }\href@noop {} {\bibfield
  {journal} {\bibinfo  {journal} {Faraday Discuss.}\ }\textbf {\bibinfo
  {volume} {184}},\ \bibinfo {pages} {173} (\bibinfo {year}
  {2015})}\BibitemShut {NoStop}%
\bibitem [{\citenamefont {Doherty}\ \emph {et~al.}(2013)\citenamefont
  {Doherty}, \citenamefont {Manson}, \citenamefont {Delaney}, \citenamefont
  {Jelezko}, \citenamefont {J.},\ and\ \citenamefont
  {Hollenberg}}]{NVcomputing2}%
  \BibitemOpen
  \bibfield  {author} {\bibinfo {author} {\bibfnamefont {M.~W.}\ \bibnamefont
  {Doherty}}, \bibinfo {author} {\bibfnamefont {N.~B.}\ \bibnamefont {Manson}},
  \bibinfo {author} {\bibfnamefont {P.}~\bibnamefont {Delaney}}, \bibinfo
  {author} {\bibfnamefont {F.}~\bibnamefont {Jelezko}}, \bibinfo {author}
  {\bibfnamefont {W.}~\bibnamefont {J.}}, \ and\ \bibinfo {author}
  {\bibfnamefont {L.~C.~L.}\ \bibnamefont {Hollenberg}},\ }\href@noop {}
  {\bibfield  {journal} {\bibinfo  {journal} {Phys. Reports}\ }\textbf
  {\bibinfo {volume} {528}},\ \bibinfo {pages} {1} (\bibinfo {year}
  {2013})}\BibitemShut {NoStop}%
\bibitem [{\citenamefont {Luo}\ \emph {et~al.}(2016)\citenamefont {Luo},
  \citenamefont {Li}, \citenamefont {Lai},\ and\ \citenamefont {Wang}}]{DOF1}%
  \BibitemOpen
  \bibfield  {author} {\bibinfo {author} {\bibfnamefont {M.-X.}\ \bibnamefont
  {Luo}}, \bibinfo {author} {\bibfnamefont {H.-R.}\ \bibnamefont {Li}},
  \bibinfo {author} {\bibfnamefont {H.}~\bibnamefont {Lai}}, \ and\ \bibinfo
  {author} {\bibfnamefont {X.}~\bibnamefont {Wang}},\ }\href@noop {} {\bibfield
   {journal} {\bibinfo  {journal} {Scientific. Reports}\ }\textbf {\bibinfo
  {volume} {6}},\ \bibinfo {pages} {1} (\bibinfo {year} {2016})}\BibitemShut
  {NoStop}%
\bibitem [{\citenamefont {Zhang}\ \emph {et~al.}(2016)\citenamefont {Zhang},
  \citenamefont {Ding}, \citenamefont {Dong}, \citenamefont {Shi},
  \citenamefont {Wang}, \citenamefont {Liu}, \citenamefont {Li}, \citenamefont
  {Zhou}, \citenamefont {Shi},\ and\ \citenamefont {Guo}}]{DOF2}%
  \BibitemOpen
  \bibfield  {author} {\bibinfo {author} {\bibfnamefont {W.}~\bibnamefont
  {Zhang}}, \bibinfo {author} {\bibfnamefont {D.-S.}\ \bibnamefont {Ding}},
  \bibinfo {author} {\bibfnamefont {M.-X.}\ \bibnamefont {Dong}}, \bibinfo
  {author} {\bibfnamefont {S.}~\bibnamefont {Shi}}, \bibinfo {author}
  {\bibfnamefont {K.}~\bibnamefont {Wang}}, \bibinfo {author} {\bibfnamefont
  {S.-L.}\ \bibnamefont {Liu}}, \bibinfo {author} {\bibfnamefont
  {Y.}~\bibnamefont {Li}}, \bibinfo {author} {\bibfnamefont {Z.-Y.}\
  \bibnamefont {Zhou}}, \bibinfo {author} {\bibfnamefont {B.-S.}\ \bibnamefont
  {Shi}}, \ and\ \bibinfo {author} {\bibfnamefont {G.}~\bibnamefont {Guo}},\
  }\href@noop {} {\bibfield  {journal} {\bibinfo  {journal} {Nature Comm.}\
  }\textbf {\bibinfo {volume} {7}},\ \bibinfo {pages} {13514} (\bibinfo {year}
  {2016})}\BibitemShut {NoStop}%
\bibitem [{\citenamefont {Wang}\ \emph {et~al.}(2015)\citenamefont {Wang},
  \citenamefont {Cai}, \citenamefont {Su}, \citenamefont {Chen}, \citenamefont
  {Wu}, \citenamefont {Li}, \citenamefont {Liu}, \citenamefont {Lu},\ and\
  \citenamefont {Pan}}]{DOF3}%
  \BibitemOpen
  \bibfield  {author} {\bibinfo {author} {\bibfnamefont {X.~L.}\ \bibnamefont
  {Wang}}, \bibinfo {author} {\bibfnamefont {X.~D.}\ \bibnamefont {Cai}},
  \bibinfo {author} {\bibfnamefont {Z.~E.}\ \bibnamefont {Su}}, \bibinfo
  {author} {\bibfnamefont {M.~C.}\ \bibnamefont {Chen}}, \bibinfo {author}
  {\bibfnamefont {D.}~\bibnamefont {Wu}}, \bibinfo {author} {\bibfnamefont
  {L.}~\bibnamefont {Li}}, \bibinfo {author} {\bibfnamefont {N.~I.}\
  \bibnamefont {Liu}}, \bibinfo {author} {\bibfnamefont {C.~Y.}\ \bibnamefont
  {Lu}}, \ and\ \bibinfo {author} {\bibfnamefont {J.~W.}\ \bibnamefont {Pan}},\
  }\href@noop {} {\bibfield  {journal} {\bibinfo  {journal} {Nature}\ }\textbf
  {\bibinfo {volume} {518}},\ \bibinfo {pages} {7540} (\bibinfo {year}
  {2015})}\BibitemShut {NoStop}%
\bibitem [{\citenamefont {Munro}\ \emph {et~al.}(2012)\citenamefont {Munro},
  \citenamefont {Stephens}, \citenamefont {Harrison},\ and\ \citenamefont
  {Nemoto}}]{DOF_Bill}%
  \BibitemOpen
  \bibfield  {author} {\bibinfo {author} {\bibfnamefont {W.~J.}\ \bibnamefont
  {Munro}}, \bibinfo {author} {\bibfnamefont {A.~M.}\ \bibnamefont {Stephens}},
  \bibinfo {author} {\bibfnamefont {K.~A.}\ \bibnamefont {Harrison}}, \ and\
  \bibinfo {author} {\bibfnamefont {K.}~\bibnamefont {Nemoto}},\ }\href@noop {}
  {\bibfield  {journal} {\bibinfo  {journal} {Nature Photonics}\ }\textbf
  {\bibinfo {volume} {6}},\ \bibinfo {pages} {777} (\bibinfo {year}
  {2012})}\BibitemShut {NoStop}%
\bibitem [{\citenamefont {Pan}\ \emph {et~al.}(2001)\citenamefont {Pan},
  \citenamefont {Simon}, \citenamefont {Brukner},\ and\ \citenamefont
  {Zeilinger}}]{Quantum_pur1}%
  \BibitemOpen
  \bibfield  {author} {\bibinfo {author} {\bibfnamefont {J.~W.}\ \bibnamefont
  {Pan}}, \bibinfo {author} {\bibfnamefont {C.}~\bibnamefont {Simon}}, \bibinfo
  {author} {\bibfnamefont {C.}~\bibnamefont {Brukner}}, \ and\ \bibinfo
  {author} {\bibfnamefont {A.}~\bibnamefont {Zeilinger}},\ }\href@noop {}
  {\bibfield  {journal} {\bibinfo  {journal} {Nature}\ ,\ \bibinfo {pages}
  {410}} (\bibinfo {year} {2001})}\BibitemShut {NoStop}%
\bibitem [{\citenamefont {Devitt}\ \emph
  {et~al.}(2013{\natexlab{b}})\citenamefont {Devitt}, \citenamefont {Munro},\
  and\ \citenamefont {Nemoto}}]{devitt}%
  \BibitemOpen
  \bibfield  {author} {\bibinfo {author} {\bibfnamefont {S.}~\bibnamefont
  {Devitt}}, \bibinfo {author} {\bibfnamefont {W.~J.}\ \bibnamefont {Munro}}, \
  and\ \bibinfo {author} {\bibfnamefont {K.}~\bibnamefont {Nemoto}},\
  }\href@noop {} {\bibfield  {journal} {\bibinfo  {journal} {Reports on
  Progress in Physics}\ }\textbf {\bibinfo {volume} {76}},\ \bibinfo {pages}
  {076001} (\bibinfo {year} {2013}{\natexlab{b}})}\BibitemShut {NoStop}%
\bibitem [{pri()}]{private}%
  \BibitemOpen
  \href@noop {} {\bibinfo  {journal} {Private Communication}\ }\BibitemShut
  {NoStop}%
\bibitem [{\citenamefont {Shor}(1997)}]{Err.corr1}%
  \BibitemOpen
\bibfield  {journal} {  }\bibfield  {author} {\bibinfo {author} {\bibfnamefont
  {P.~W.}\ \bibnamefont {Shor}},\ }\href@noop {} {\bibfield  {journal}
  {\bibinfo  {journal} {SIAM Journal of Sci. Statist. Comput.}\ }\textbf
  {\bibinfo {volume} {26}},\ \bibinfo {pages} {1484} (\bibinfo {year}
  {1997})}\BibitemShut {NoStop}%
\bibitem [{\citenamefont {Bennett}\ \emph
  {et~al.}(1996{\natexlab{c}})\citenamefont {Bennett}, \citenamefont
  {DiVincenzo}, \citenamefont {Smolin},\ and\ \citenamefont
  {Wooters}}]{Err.corr2}%
  \BibitemOpen
  \bibfield  {author} {\bibinfo {author} {\bibfnamefont {C.~H.}\ \bibnamefont
  {Bennett}}, \bibinfo {author} {\bibfnamefont {D.~P.}\ \bibnamefont
  {DiVincenzo}}, \bibinfo {author} {\bibfnamefont {J.~A.}\ \bibnamefont
  {Smolin}}, \ and\ \bibinfo {author} {\bibfnamefont {W.~K.}\ \bibnamefont
  {Wooters}},\ }\href@noop {} {\bibfield  {journal} {\bibinfo  {journal} {Phys.
  Rev. A}\ }\textbf {\bibinfo {volume} {54}},\ \bibinfo {pages} {3824}
  (\bibinfo {year} {1996}{\natexlab{c}})}\BibitemShut {NoStop}%
\bibitem [{\citenamefont {Gottesman}(2000)}]{Err.corr3}%
  \BibitemOpen
  \bibfield  {author} {\bibinfo {author} {\bibfnamefont {D.}~\bibnamefont
  {Gottesman}},\ }\href@noop {} {\bibfield  {journal} {\bibinfo  {journal} {J.
  Mod. Opt.}\ }\textbf {\bibinfo {volume} {47}},\ \bibinfo {pages} {333}
  (\bibinfo {year} {2000})}\BibitemShut {NoStop}%
\bibitem [{\citenamefont {D.}\ \emph {et~al.}(2005)\citenamefont {D.},
  \citenamefont {Laflamme},\ and\ \citenamefont {Poulin}}]{Err.corr4}%
  \BibitemOpen
  \bibfield  {author} {\bibinfo {author} {\bibfnamefont {K.}~\bibnamefont
  {D.}}, \bibinfo {author} {\bibfnamefont {R.}~\bibnamefont {Laflamme}}, \ and\
  \bibinfo {author} {\bibfnamefont {D.}~\bibnamefont {Poulin}},\ }\href@noop {}
  {\bibfield  {journal} {\bibinfo  {journal} {Phys. Rev. Lett.}\ }\textbf
  {\bibinfo {volume} {94}},\ \bibinfo {pages} {180501} (\bibinfo {year}
  {2005})}\BibitemShut {NoStop}%
\bibitem [{\citenamefont {Fowler}\ \emph {et~al.}(2009)\citenamefont {Fowler},
  \citenamefont {Stephens},\ and\ \citenamefont {Groszkowski}}]{Err.corr5}%
  \BibitemOpen
  \bibfield  {author} {\bibinfo {author} {\bibfnamefont {A.~G.}\ \bibnamefont
  {Fowler}}, \bibinfo {author} {\bibfnamefont {A.~M.}\ \bibnamefont
  {Stephens}}, \ and\ \bibinfo {author} {\bibnamefont {Groszkowski}},\
  }\href@noop {} {\bibfield  {journal} {\bibinfo  {journal} {Phys. Rev. A}\
  }\textbf {\bibinfo {volume} {80}},\ \bibinfo {pages} {052312} (\bibinfo
  {year} {2009})}\BibitemShut {NoStop}%
\bibitem [{\citenamefont {Knill}(2005{\natexlab{b}})}]{Err.corr6}%
  \BibitemOpen
  \bibfield  {author} {\bibinfo {author} {\bibfnamefont {E.}~\bibnamefont
  {Knill}},\ }\href@noop {} {\bibfield  {journal} {\bibinfo  {journal} {Nature
  (London)}\ }\textbf {\bibinfo {volume} {434}},\ \bibinfo {pages} {39}
  (\bibinfo {year} {2005}{\natexlab{b}})}\BibitemShut {NoStop}%
\bibitem [{\citenamefont {Bombin}(2011)}]{Err.corr7}%
  \BibitemOpen
  \bibfield  {author} {\bibinfo {author} {\bibfnamefont {H.}~\bibnamefont
  {Bombin}},\ }\href@noop {} {\bibfield  {journal} {\bibinfo  {journal} {New.
  J. Phys.}\ }\textbf {\bibinfo {volume} {13}},\ \bibinfo {pages} {043005}
  (\bibinfo {year} {2011})}\BibitemShut {NoStop}%
\bibitem [{\citenamefont {Bennett}\ and\ \citenamefont
  {Brassard}(1984)}]{Prepare_meas_prot}%
  \BibitemOpen
  \bibfield  {author} {\bibinfo {author} {\bibfnamefont {C.~H.}\ \bibnamefont
  {Bennett}}\ and\ \bibinfo {author} {\bibfnamefont {G.}~\bibnamefont
  {Brassard}},\ }\href@noop {} {\bibfield  {journal} {\bibinfo  {journal}
  {Quantum cryptography: Public key distribution and coin tossing. In:
  Proceedings of IEEE International Conference on Computers, Systems, and
  Signal Processing}\ ,\ \bibinfo {pages} {175}} (\bibinfo {year}
  {1984})}\BibitemShut {NoStop}%
\bibitem [{\citenamefont {Sangouard}\ \emph
  {et~al.}(2011{\natexlab{b}})\citenamefont {Sangouard}, \citenamefont {Simon},
  \citenamefont {de~Riedmatten},\ and\ \citenamefont {Gisin}}]{P0}%
  \BibitemOpen
  \bibfield  {author} {\bibinfo {author} {\bibfnamefont {N.}~\bibnamefont
  {Sangouard}}, \bibinfo {author} {\bibfnamefont {C.}~\bibnamefont {Simon}},
  \bibinfo {author} {\bibfnamefont {H.}~\bibnamefont {de~Riedmatten}}, \ and\
  \bibinfo {author} {\bibfnamefont {N.}~\bibnamefont {Gisin}},\ }\href@noop {}
  {\bibfield  {journal} {\bibinfo  {journal} {Review of Modern Physics}\
  }\textbf {\bibinfo {volume} {83}} (\bibinfo {year}
  {2011}{\natexlab{b}})}\BibitemShut {NoStop}%
\end{thebibliography}%

\end{document}